\documentclass[12pt]{iopart}
\usepackage{graphicx,amssymb}  

\def\gs{\mathrel{\raise1.16pt\hbox{$>$}\kern-7.0pt\lower3.06pt\hbox{{$\scriptstyle \sim$}}}}
\def\ls{\mathrel{\raise1.16pt\hbox{$<$}\kern-7.0pt\lower3.06pt\hbox{{$\scriptstyle \sim$}}}}

\begin{document}

\title[The dark matter of gravitational lensing]{The dark matter of gravitational lensing}

\author{Richard Massey, Thomas Kitching}
\address{Institute for Astronomy, Royal Observatory, Blackford Hill, Edinburgh EH9 3HJ, UK}
\ead{rm,tdk@roe.ac.uk}
\author{Johan Richard}
\address{Durham University, Department of Physics, South Road, Durham DH1 3LE, UK}
\ead{johan.richard@durham.ac.uk}
\begin{abstract}

We review progress in understanding dark matter by astrophysics, and particularly via the effect of
gravitational lensing. Evidence from many different directions now all imply that five 
sixths of the material content of
the universe is in this mysterious form, separate from and beyond the ordinary ``baryonic'' particles in the
standard model of particle physics. Dark matter appears not to interact via the electromagnetic 
force, and therefore neither
emits nor reflects light. However, it definitely does interact via gravity, and has played the most important
role in shaping the Universe on large scales. The most successful technique with which to investigate it has
so far been the effect of gravitational lensing. The curvature of space-time near any gravitating mass
(including dark matter) deflects passing rays of light -- observably shifting, distorting and magnifying the
images of background galaxies. Measurements of such effects currently provide constraints on the mean density
of dark matter, and its density relative to baryonic matter; the size and mass of individual dark matter
particles; and its cross section under various fundamental forces.

\end{abstract}

\submitto{\RPP}
\maketitle

\section{Introduction}

Astrophysics now operates under the astonishing hypothesis that the Universe we see is but the tip of an iceberg.
It has taken a wealth of evidence from many independent observations to confirm that, while the ``standard model'' of 
particle physics may successfully describe quarks, leptons and bosons, it misses the most common form of matter.
The first evidence for this provocative stance came from the unexpectedly
high velocities of galaxies in the Coma cluster \cite{zwicky33,zwicky37} and Virgo cluster \cite{smith36}. 
The clusters appear to be gravitationally bound, but all the luminous material inside them
does not add up to sufficient mass to retain the fast-moving galaxies.
In individual galaxies too, stars orbit too fast to be held by the luminous material of
Andromeda \cite{babcock39,rubin70,roberts75}, NGC3115 \cite{oort40} and other spiral galaxies \cite{rubin85,persic88}. 
Luminous material in these galaxies is concentrated in the central regions, so the angular rotation of stars ought to
slow at large radii, but stars in the outskirts are seen to rotate at the same rate as those near the centre. Given the high 
velocities of their constituents, both galaxies and clusters of galaxies ought to pull themselves apart. 
Preserving these self-destructive systems requires gravitational glue in the form of invisible ``dark  matter''.

The modern ``concordance'' cosmological model also relies upon the gravitational influence of (cold) dark matter to glue
together the entire Universe. Slowing the expansion after the Big Bang required much more gravity than that provided by
the baryons, which alone would have allowed the contents of the Universe to be spread unhabitably thin \cite{coc04}.
However, if additional dark matter were forged during the same primordial fireball, it must have quickly stopped
interacting with other particles through the electroweak force, in order to preserve the uniformity of photons in the
Cosmic Microwave Background (CMB) radiation, whose temperature fluctuations reach only one part in $10^{5}$ of the mean \cite{peebles82}.
Indeed, once dark matter decoupled from standard model particles, it began collapsing under its own gravity into dense concentrations of
mass. These provided the initial scaffolding for structure formation.
Once ordinary matter had cooled further, and also decoupled from the hot photons, it could fall into the scaffolding and be
built into galaxies \cite{davis85,efstathiou85,efstathiou90,springel06}. Without dark matter's headstart, there would
have been insufficient time to build the complex structures we see (and live in) today. The latest measurements of the CMB and 
Large-Scale Structure \cite{dunkley09,lesgourgues07} indicate that the Universe contains
approximately one hydrogen atom per cubic metre, but five times that in the form of dark matter.  

On the back of this evidence, determining the nature of ubiquitous dark matter has become an outstanding problem of modern
physics. Yet its low rate of interaction with the rest of the Universe makes it difficult to detect.
Since dark matter does not generally emit, reflect or absorb light of any wavelength, traditional astrophysics is rendered blind. 
No particle colliders have yet achieved sufficient energy to create a single dark matter particle
and, even if they can identify a dark matter candidate,
\textit{astronomical observations will still be required to demonstrate that the candidate particle is present in
sufficient quantities throughout the Universe to be the dark matter}. 
Direct detection experiments in quiet, underground mines have yet to locate a convincing signal -- and explaining this
absence also requires astrophysical explanations, such as the patchiness of dark matter debris from consumed satellite 
galaxies \cite{chapman08,cole08}. 
Astronomy will therefore remain vital in resolving the outstanding problem that it initiated.

As with the first detections, the best way to study dark matter is via its gravitational influence on more easily
visible particles. The most direct method for this is ``gravitational lensing'', the deflection of photons as they
pass through the warped space-time of a gravitational field \cite{einstein16}. Light rays from distant sources are
not ``straight'' (in a Euclidean frame) if they pass near massive objects, such as stars, clusters of galaxies or
dark matter, along our line of sight. In practice, the effect is similar to optical refraction, although it arises
from very different physics. 
The effect was first observed in $1919$, during a solar eclipse in front of the Hyades star cluster,
whose stars appeared to move as they passed behind the mass of the sun \cite{dyson19}.
This observation provided the first experimental
verification of general relativity. Although neither Einstein nor the observers saw any further uses for the
effect \cite{einstein36}, Zwicky suggested that the ultimate measurement of cluster masses 
would come from lensing \cite{zwicky37}, and it has indeed become the most successful probe of the dark sector. 

Many lines of research currently exploit the effect of gravitational lensing. It is a rapidly growing field, but the
first threads of consensus are beginning to emerge in answer to the top-level questions. As we shall discuss in this
review, the technique has provided vital contributions to the following deductions:


\begin{itemize}
\item The Universe contains about five times more dark matter than baryonic matter
\item Dark matter interacts approximately normally via gravity 
\item Dark matter has a very small electroweak and self-interaction cross section
\item Dark matter is not in the form of dense, planet-sized bodies
\item Dark matter is dynamically cold.
\end{itemize}

We introduce the various observational flavours of gravitational lensing in \S\ref{sec:progress}. We then describe
lensing measurements that have shed light upon the amount of dark matter in \S\ref{sec:quantity}, its organisation in
\S\ref{sec:organisation}, and its properties in \S\ref{sec:properties}. We try to touch upon all of the areas of
gravitational lensing that have contributed to current knowledge of dark matter, but cannot comprehensively discuss
observations from all fields in a single review. We discuss future prospects and challenges for the field in
\S\ref{sec:future}, and conclude in \S\ref{sec:conclusions}. Note that measurements of cosmological distances all depend
upon the overall rate of expansion of the Universe, parameterised as Hubble's constant $H$. Throughout this review, we
assume a background cosmological model in which this value at the present day is $h=H_0/^0=70$~km/s/Mpc. Since this is
only known to $\sim5$\% accuracy \cite{riess09}, uncertainty in the geometry of gravitational lens systems propagates
implicitly into the same uncertainty in all inferred (absolute) lens masses.

\section{Observational flavours of Gravitational Lensing}\label{sec:progress}

\subsection{Strong lensing}

\begin{figure}[!t]
\begin{flushright}
\includegraphics[width = 372pt]{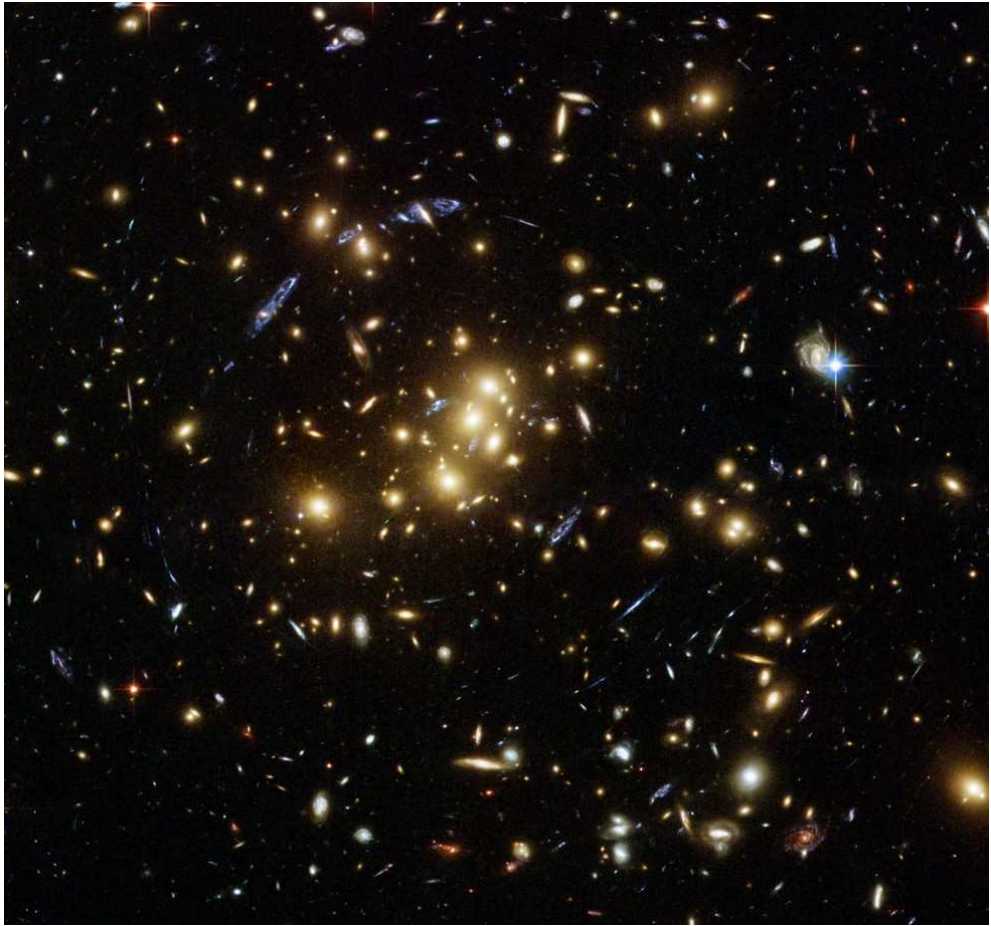}
\end{flushright}
\caption{\footnotesize{
Strong gravitational lensing around galaxy cluster CL0024+17, demonstrating at least
three layers projected onto a single  2D image. The + shaped objects are nearby stars
in our own galaxy (the + created by optical effects in the telescope).
The yellow, elliptical galaxies are members of the cluster, all at a similar redshift and
gravitationally bound. Also amongst this group  of galaxies is a halo of invisible dark
matter. The elongated blue objects are much more distant galaxies, physically unassociated with, and
lying behind, the cluster. Gravitational lensing has distorted their apparent images
into a series of tangential arcs. Figure credit: NASA/ESA/M.J. Jee (John Hopkins
University).}}\label{fig:strongdemo}
\end{figure}

Gravitational lensing is most easily observable around a dense concentration of mass like the core of a galaxy or
cluster of galaxies. In the ``strong lensing'' regime, nearby space-time is so warped that light can travel along
multiple paths around the lens, and still be deflected back towards the observer \cite{treurev}. If a distant
source is directly behind a circular lens, the light can travel around any side of it, and appears as an ``Einstein
ring''. The Einstein radius or size of this ring is proportional to the square root of the projected mass inside it. If the background
source is slightly offset, or the lens has a complex shape, the source can still appear in multiple locations,
viewed from very slightly different angles. Depending on the focussing of the light path, each of these multiple
images can be made brighter (magnified) or fainter (demagnified), and the magnification is greatest close to the
``critical curve'' (the asymmetric equivalent of an Einstein ring) \cite{breen87}. Since light from opposite ends of an extended
source (e.g.\ a galaxy) is typically deflected by different amounts, the source appears distorted. Distant galaxies 
intrinsically no different from any others appear as tangential
arcs around the lens or, if the lens mass is very concentrated, a line radiating away from it \cite{hammer97}. Such ``radial arcs''
are generally difficult to see because they are usually less magnified and appear inside the Einstein 
radius, behind any light emitted by the lens object itself. 
An example of strong gravitational lensing around a massive galaxy cluster is shown in figure~\ref{fig:strongdemo}.

The first strong gravitational lens was discovered with the Jodrell Bank MkIA radio telescope in 1979 \cite{walsh}.
Two quasars were found $6$ arcseconds apart, with identical redshifts $z=1.41$ and detailed absorption spectra. A
foreground $z=0.355$ galaxy is now known between them. Subsequent observational progress has then driven primarily
by technological advance. Photographic plates even on large telescopes and scanned by computers gather light too
inefficiently to capture optical images of the distant (therefore faint) and thin lensed arcs. Digital CCD cameras
have far higher efficiency, and the first image of a strongly lensed arc was obtained with the Canada-France Hawaii
telescope in galaxy cluster Abell 370 \cite{soucail87,kovner87}, and confirmed to be a single object at redshift $z=0.72$ by
optical spectroscopy \cite{soucail88}. Several more such giant arcs were quickly identified in other galaxy
clusters \cite{lynds89}, soon the first sample of strong lensing clusters was built \cite{lefevre94} and allowed 
for statistical study of giant arcs \cite{wu93}.

The launch of the Hubble Space Telescope (HST) then revolutionised the field once again. Its unrivalled imaging
resolution helped distinguish a large number of arcs, arclets and multiple images in many clusters. The first study
using the Wide Field Planetary Camera (WFPC2) identified seven strongly lensed objects behind the cluster Abell~2218 \cite{kneib96},
significantly more than bad been found by using ground-based telescopes. The Advanced Camera for Surveys (ACS)
provided a further step forward, with $20$-$30$ strongly lensed objects found in several of the most massive clusters \cite{broadhurst05}
and over $100$ multiple images around Abell~1689 \cite{limousin07}. The positions and shapes of the images can be
used to reconstruct the distribution of mass in the lens. The magnification effect boosts the observed fluxes of 
background objects, so that a strong lensing cluster can also be used as a
\textit{gravitational telescope} to see -- and even resolve -- fainter or more distant objects 
than otherwise possible \cite{starck08}.

The detection of strong-lensing events on galaxy scales also enabled constraints on cosmological parameters using 
large statistical samples. Throughout the 1990s, the Cosmic Lens All-Sky Survey (CLASS, \cite{myers03}) searched for 
gravitationally lensed compact radio sources using imaging from the Very Large Array (VLA). Out of 
$\sim$16,500 radio sources, they found 22 lens systems. The statistical properties of these lensed systems constrained 
cosmological parameters \cite{chae03} and measurements of their time delays constrained Hubble's constant \cite{koopmans01}. 
More recently, systematic homogeneous surveys such as the Sloan Digital Sky Survey (SDSS) have provided even larger samples of 
strong-lensing galaxies. The SDSS Quasar Lens Search (SQLS, \cite{oguri06}) spectroscopically found 53 lensing galaxies
and tightened constraints on cosmological parameters \cite{oguri08} by ingeniously looking for the signature of two objects 
at different redshifts. Finally, the Sloan Lens ACS (SLACS, \cite{bolton06}) Survey combined the massive 
data volume of SDSS with the high-resolution imaging capability of ACS to identify and then follow up 131 galaxy-galaxy lensing systems 
\cite{bolton08}, measuring the average dark matter fraction and dark matter density profiles within galaxies \cite{gavazzi07}.

\subsection{Microlensing}

Most distant astronomical observations are static on the scale of a human lifetime but, as in the case of the 1919
eclipse, an exception is provided by any relative motion between a source and a gravitational lens. The line of
sight to a star along which a foreground mass would induce gravitational lensing represents a tiny volume of space.
Panoramic imaging cameras now make it possible to monitor the lines of sight to many millions of stars, and any
object traversing any of those small volumes can temporarily brighten it for days or weeks. Indeed, ``pixel lensing'' of even
unresolved stars can still detect the statistical passage
of a foreground lens in front of one of the many stars contributing to the light in any pixel \cite{gould96,novati09b}.
The main observational concern is to avoid false-positive detections due to intrinsic variability in the luminosity
of certain types of star. The two most exciting results microlensing studies are that dark matter in the
Milky Way is not predominantly in the form of freefloating, planet-sized lumps of dull rock, which would occasionally brighten stars in 
the Galactic centre, but that planet-sized lumps of rock do exist around other stars, and give rise to secondary brighness peaks shortly
before or after their host star itself acts as a gravitational lens.

The term ``gravitational microlensing'' was coined by Refsdal \cite{refsdal64,paczynski86}, from the characteristic $\sim1$~microarcsecond
size of a star's Einstein radius. Only physically small sources will be significantly affected by microlensing; extended background
sources like galaxies are effectively immune because only a tiny fraction of their light is strongly magnified, with the rest propagating
unaffected.  More massive lenses, with milliarcsecond Einstein radii, produce ``gravitational millilensing'' that affects slightly larger
background sources on a timescale of months (and the statistical long tail is strong lensing around massive clusters, with arcsecond
Einstein radii). This distinction has been most useful when looking at the lensed images of Active Galactic Nuclei (a galaxy's central,
supermassive black hole and surrounding accretion disc), because these really do have different physical sizes when viewed at different
wavelengths. As matter gradually falls into the black hole, it emits a warm glow of infra-red light from the large and outer narrow-line
region, then optical light from the smaller broad-line region and finally ultra-violet light from the accretion disc itself. The behaviour
of the source can be modelled from long wavelength observations, which are relatively unaffected by gravitational lensing, then the lens
object and even its substructure probed at progressively shorter wavelengths.

\subsection{Weak lensing}

Most lines of sight through the Universe do not pass near a strong gravitational lens. Far from the core of a
galaxy or cluster of galaxies, the light deflection is very slight. In this ``weak lensing'' regime,  the
distortion of resolved sources can be approximated to first order as a locally linear transformation of the sky,
represented as a $2\times 2$ matrix that includes magnification, shear and (potentially, but not usually in
practice) rotation \cite{mellier99,bartelmann99,refregier04,munshi07,hoekstrarev}. 
The theory was developed during the 1990s
\cite{blandford,kaiser}, including some practical methods to accurately measure galaxy positions and shapes in the
new pixellated CCD images \cite{ksb95}. Either the magnification \cite{menard09} or the shear distortion can be measured, but the
shear tends to have higher signal to noise, because competing effects of magnification (the brightening of faint
galaxies, but the dilution of the surveyed volume in a fixed angle on the sky) act against each other and partially
cancel \cite{gray}. 

The shear distortion changes the shapes of distant galaxies, adjusting their major-to-minor axis ratio by $\sim2\%$.
This cannot be seen in an individual object, since it is far smaller than the range of intrinsic shape variation 
in galaxies, which are already elliptical, have spiral arms and
knots of star formation, etc. However, galaxies along adjacent lines of sight are coherently sheared by a similar
amount, while their intrinsic shapes are (to first order) uncorrelated. In the absence of lensing, if there is no
preferred direction in the Universe, galaxy shapes must average out as circular. Once sheared, the average shape of
adjacent galaxies is an ellipse, from which the shear signal can be measured statistically. The intrinsic shapes of
galaxies are noise in this measurement (averaging over $\sim100$ galaxies is required to obtain a signal to noise
of unity in shear). The spatial resolution of this measurement is determined by the density on the sky of galaxies
whose shapes are resolved: typically a few square arcminutes from the ground, or one square arcminute from space.

\begin{figure}[!t]
\begin{flushright}
\includegraphics[scale = 0.34]{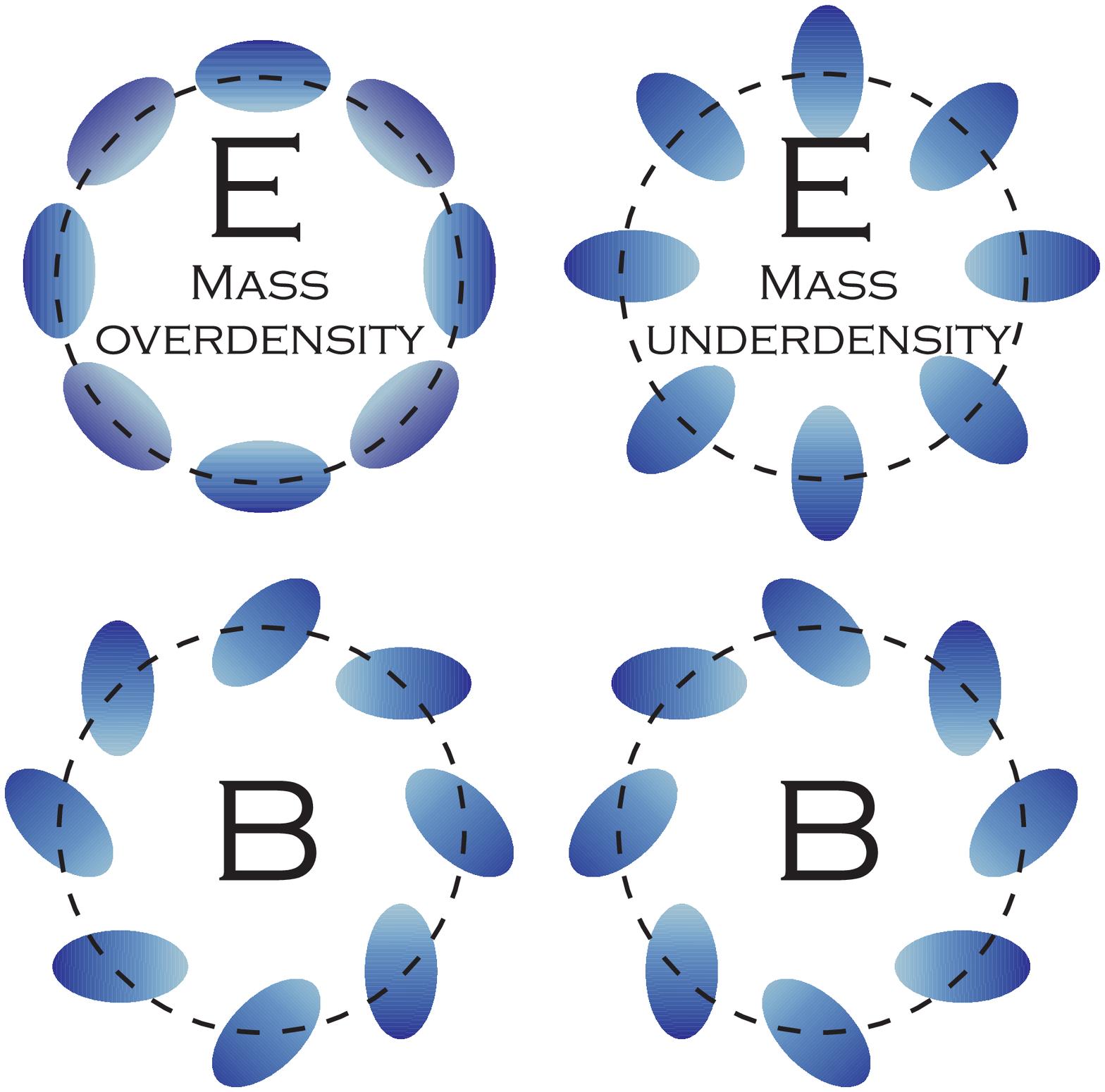}~~~~
\includegraphics[scale = 0.32]{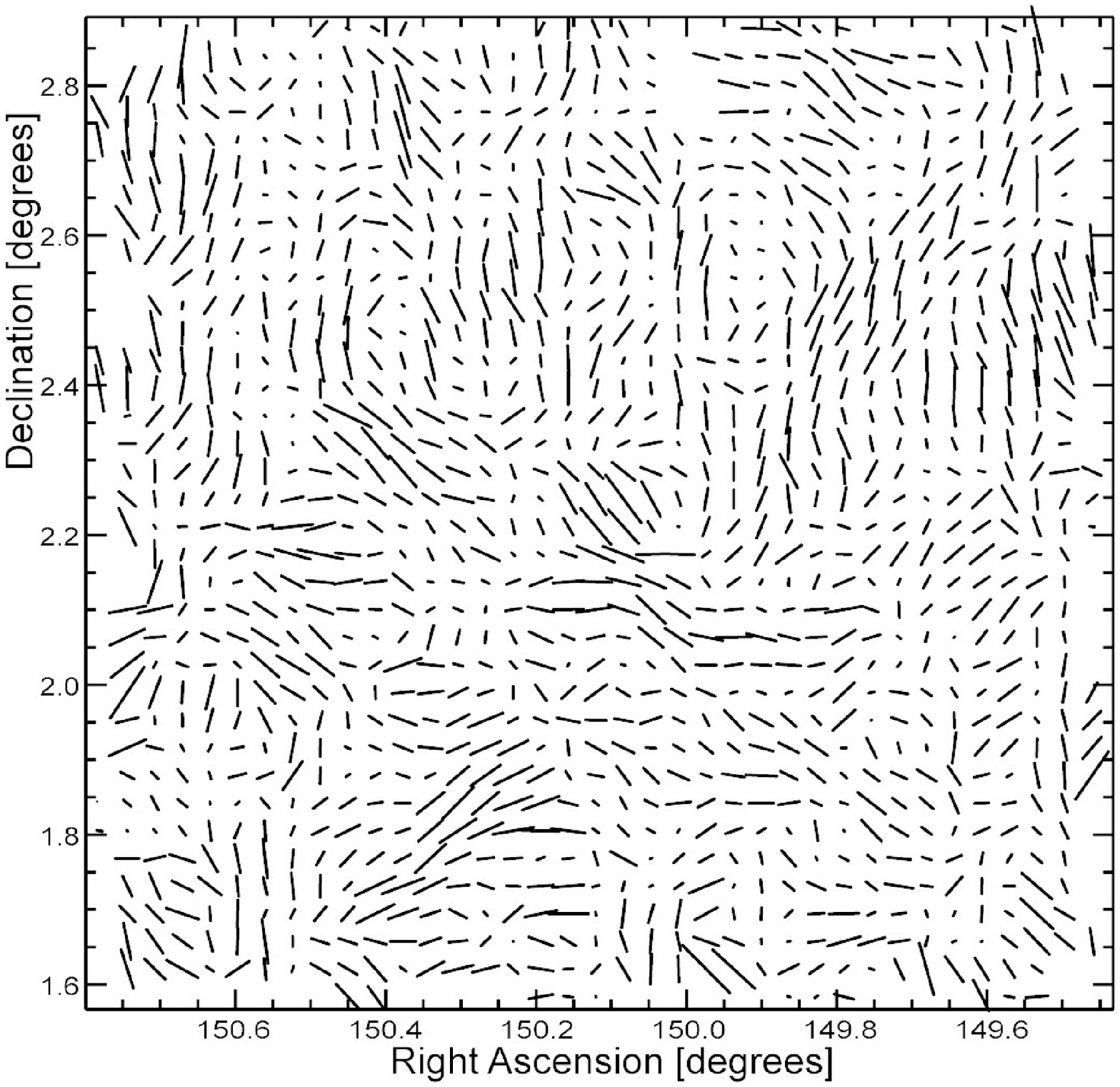}
\end{flushright}
\caption{
The statistical signals sought by measurements of weak gravitational lensing are
slight but coherent distortions in the shapes of distant galaxies.
{\it (Left):} A tangential, circular
pattern of background galaxies is produced around a foreground mass overdensity, reminiscent of the tangential arcs
of strong lensing seen in figure~\ref{fig:strongdemo}. On much larger scales, an  opposite, radial pattern is
produced by foreground voids. Physical gravitational lensing produces  only these ``$E$-mode''
patterns. However, there is another degree of freedom in a shear (vector-like) field, and
spurious artefacts can typically mimic both. Measurements of ``$B$-mode'' patterns 
therefore provide a free test for residual systematic defects.
{\it (Right):} The observed ellipticities of half a million distant galaxies
within the $2$ square degree Hubble Space Telescope COSMOS survey \cite{massey07a}. Each tick mark
represents the mean ellipticity of several hundred galaxies. A dot represents a circular mean 
galaxy; lines represent elliptical mean galaxies, with
the length of the line proportional to the ellipticity, and in the direction of the major axis.
The longest lines represent an ellipticity of about $0.06$. Several coherent circular $E-$mode
patterns are evident in this figure, e.g.\ ($149.9$, $2.5$). Radial $E-$mode
patterns are also present on larger scales, but the density in voids cannot be negative,
so the contrast is lower and the signal much less apparent to the eye. 
The $B-$mode signal is consistent with zero.}\label{fig:eb}
\end{figure}

The observable shear field is proportional to a second derivative of the gravitational potential projected along a
line of sight. Via a convolution, this can be converted into a map of the projected mass distribution at the same resolution. The mass
is just a different second derivative of the gravitational potential and is responsible for the circular ``$E$-mode'' patterns shown in figure~\ref{fig:eb} and reminiscent of the tangential
strong lensing arcs around clusters.
Conveniently, a second scalar quantity can be extracted from the shear field. The curl-like ``$B$-mode'' signal is not 
produced by
the gravitational field of a single mass, and only in very low amounts by a complex distribution of mass
\cite{schneider}. However, many potential systematics produce $E$ and $B$-modes equally, so checking that the 
$B$-mode is
consistent with zero in a final analysis provides a useful test that an analysis has successully removed any
residual instrumental systematics.

By the late 1990s, weak gravitational lensing had been detected around the most massive clusters, and an
optimistic outlook was presented in an influential review in 1997 \cite{mellier99}. The optimism was well
founded, for weak lensing really burst onto the cosmological scene 
during a single month in 2000 when the first large-format CCDs
allowed four groups to independently detect weak lensing in random patches of the sky
\cite{bacon00, kaiser00, vanwaerbeke00, wittman00}, a probe of the true average
distribution of dark matter. In particular, it was the consistency 
between the four independent measurements
that assuaged doubts from an initially skeptical astronomical community and laid the foundations for larger,
dedicated surveys from telescopes both on the ground and in space. Weak lensing has rapidly become a standard
cosmological tool.

\subsection{Flexion}

\begin{figure}[!t]
\begin{flushright}
\includegraphics[width = 372pt]{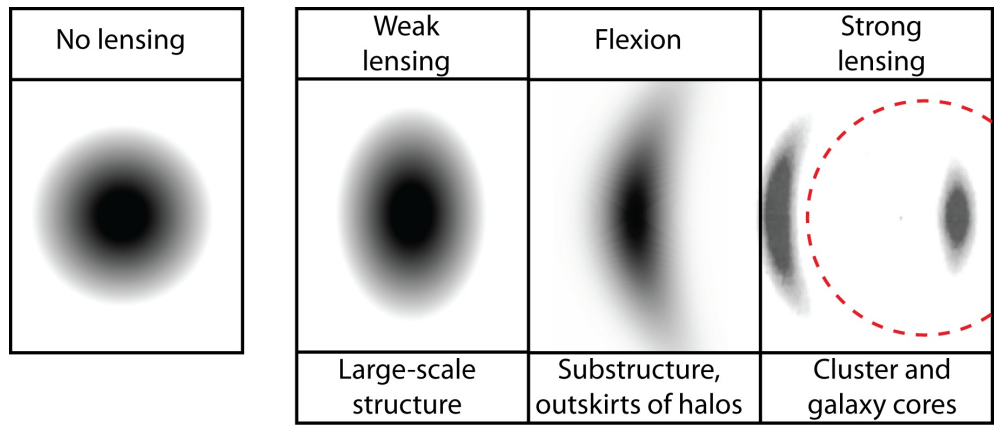}
\end{flushright}
\caption{The various regimes of gravitational lensing image distortion. 
Along typical lines of sight through the Universe, an intrinsically circular source is distorted into an ellipse by
weak lensing \textit{shear}. The resulting axis ratio is typically only $\sim2\%$ and has been exaggerated in this
figure for illustration. Nearer concentrations of mass, the distortion begins to introduce \textit{flexion}
curvature. Along lines of sight passing near the most massive galaxies of clusters of galaxies, and through the
most curved space-time, strong gravitational lensing produces multiple imaging and giant arcs.}
\label{fig:LensingRegimes} \end{figure}

Bridging the gap between strong and weak lensing is the second-order effect known as flexion. If the projected
mass distribution of a lens has a spatial gradient, steep enough to change the induced shear from one side of a
source galaxy to the other, that galaxy begins to curve as shown in figure~\ref{fig:LensingRegimes}. This is
the next term in a lensing expansion that leads towards the formation of an arc, as in strong lensing. The
amplitude of the flexion signal is lower than the shear signal, 
but so is the intrinsic curvature of typical galaxy shapes. 
Statistical techniques similar to those used in weak lensing can therefore be applied. 
Flexion measurements have proven most useful to fill in a gap in the reconstructed mass around galaxy
clusters where the light deflection is too small for strong lensing, but the area 
(and hence the number of lensed sources) is too low for a significant weak lensing analysis.

The initial attempts to mathematically describe the flexion distortion were forbidding
\cite{goldberg02,irwin03,irwin06}. More recent descriptions adapt the complex notation from weak lensing shear
into an elegant formalism requiring just one extra derivative of the gravitational potential
\cite{goldberg05,bacon06}. Flexion has an equivalent of the $E$- and $B$-mode decomposition
\cite{bernstein09}, and one extra degree of freedom in the second-order equations produces an additional distortion that is not
produced by gravitational lensing \cite{bacon09b}. Measurements of all these extra patterns may provide useful
crosschecks for residual image processing systematics.

\section{Quantity of Dark Matter} \label{sec:quantity}

\subsection{Amount of dark matter in individual galaxies}\label{sec:quantity_gals}

Individual galaxies are built of baryonic material encased inside a much larger halo of dark matter. 
Gravitational lensing can probe this halo at outer radii far beyond any visible tracers of mass. 
Indeed, there is now better agreement about the profile of the dark matter halo than the distribution of the 
central baryons!

The weak lensing signal in SDSS survey imaging is very noisy, but stacking the signal around a third of a million
galaxies reveals a typical halo of total (weak lensing) mass $1.4\times10^{12}M_\odot$ around galaxies with a stellar
mass of $6\times10^{10}M_\odot$ (as determined from a comparison of the spectrum of emitted light against theoretical
models), independently of their visual morphology \cite{mandelbaum06b}. Across all galaxies, these stars would account
for $\sim16\%$ of the expected baryons in the Universe. Within rather uncertain errors, this is consistent with
independent radio observations of atomic gas that indicate that while only $\sim10\%$ of baryons end up in galaxies,
almost all of these form stars \cite{read05b}.

To more directly measure the mass of the central baryons, the Hubble Space Telescope SLACS survey of elliptical galaxies
probes the distribution of total mass throughout galaxies by combining weak lensing with strong lensing and
parameterizing the density of dark matter. Such observations necessarily require more massive galaxies, and find haloes
of $1.2\pm0.3\times10^{13}M_\odot$ around ellipticals with stellar mass $2.6\pm0.3\times10^{11}M_\odot$
\cite{gavazzi07}. Crucially, baryons dominate the core by an order of magnitude excess over dark matter, comprise
$27\pm4\%$ of the mass in the central $\sim5$~kpc, and this fraction falls as expected to recover the constant value
consistent with cosmological measurements in the outskirts.

On the contrary, the Red-Sequence Cluster survey finds that elliptical galaxies live inside $\sim2\times$ more massive
dark matter haloes than spiral galaxies with the same stellar mass \cite{hoekstra05}. The ratio of total baryons to dark matter in bound systems is probably constant so, if these variations are real, they are most likely due to variations in the efficiencies of star formation between morphological types of galaxies. Other studies do find this to vary by the required factor of $\sim2$ \cite{guzik02} -- although this involves several assumptions about the loss of baryons from galaxies and the relative production of bright stars versus faint stars.



As shown in figure~\ref{fig:models_123}, the conversion of stars into baryons is most efficient today in galaxies 
of a characteristic mass of $10^{11}$--$10^{12}~M_\odot$ \cite{vandenbosch03,parker05,shankar06}. This scale has generally 
grown over cosmic history, although evidence is also emerging for ``cosmic downsizing'', by which activity may be 
shifting back to less massive structures \cite{juneau05,bundy09}. Either side of this scale, star formation is 
quenched by astrophysical effects, and the amount of total mass needed to support a given
luminosity increases \cite{cole01,marioni02,benson03,natarajan08a}. Even slightly smaller $\ls 10^{10}~M_\odot$
dark matter haloes form very few stars, because their shallow gravitational potential can not
gather a sufficient density of baryons that are being continually re-heated by a background of 
photoionising radiation from distant stars and quasars \cite{efstathiou92,bullock00}, or kept from being stirred and
expelled by winds and supernova explosions in any first stars \cite{vandenbosch03}. The situation is 
less clear in more massive haloes, although outflows from central supermassive
black holes certainly contribute to an inability of baryons to cool and condense into sufficiently dense regions to
then collapse into stars \cite{sanders08}.

%
%
%
%
%
%
%
%
%
%

\subsection{Amount of dark matter in groups and clusters of galaxies}

\begin{figure}[!t]
\begin{flushright}
\includegraphics[width = 170pt]{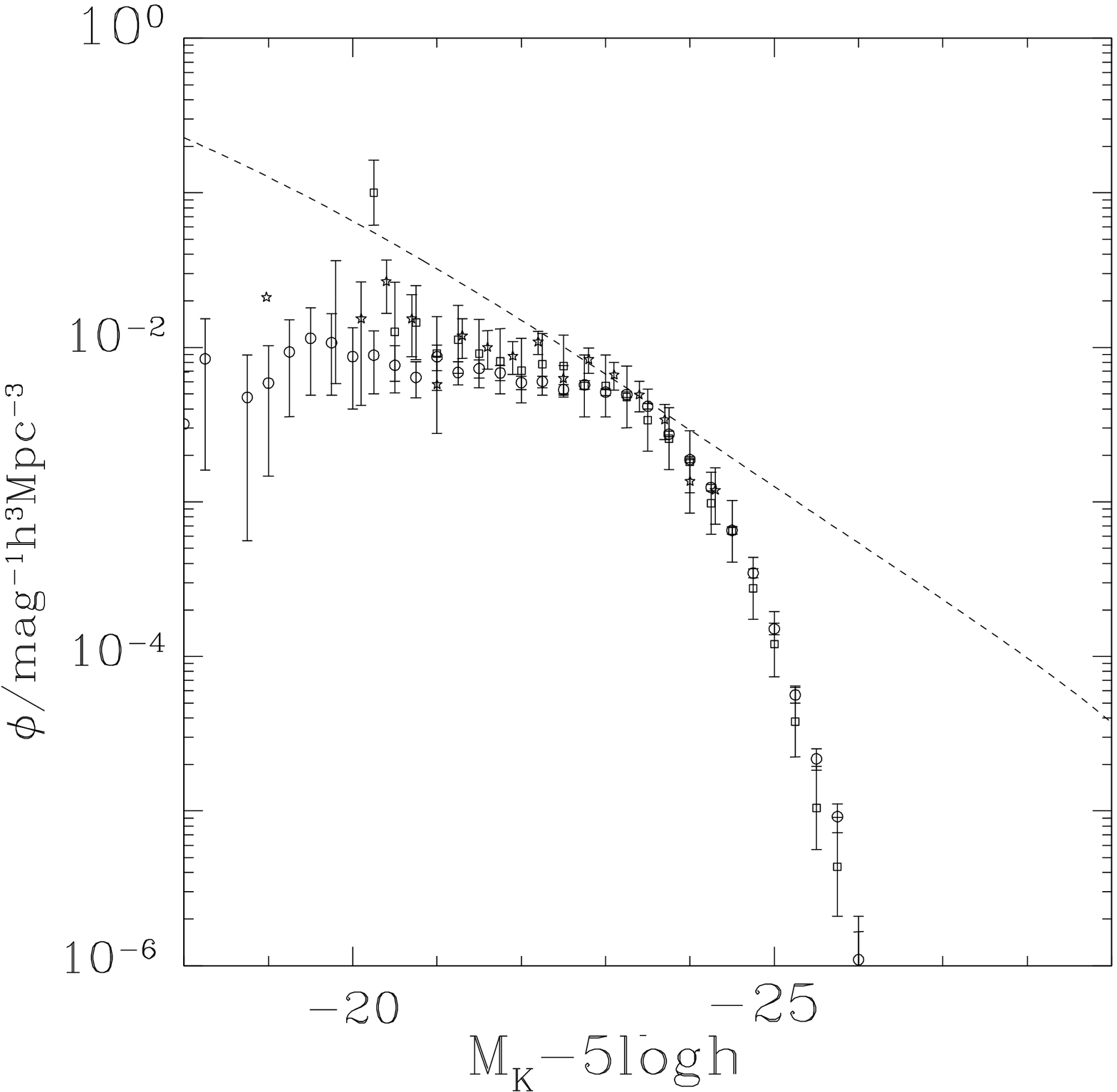}
\includegraphics[width = 185pt,angle=270]{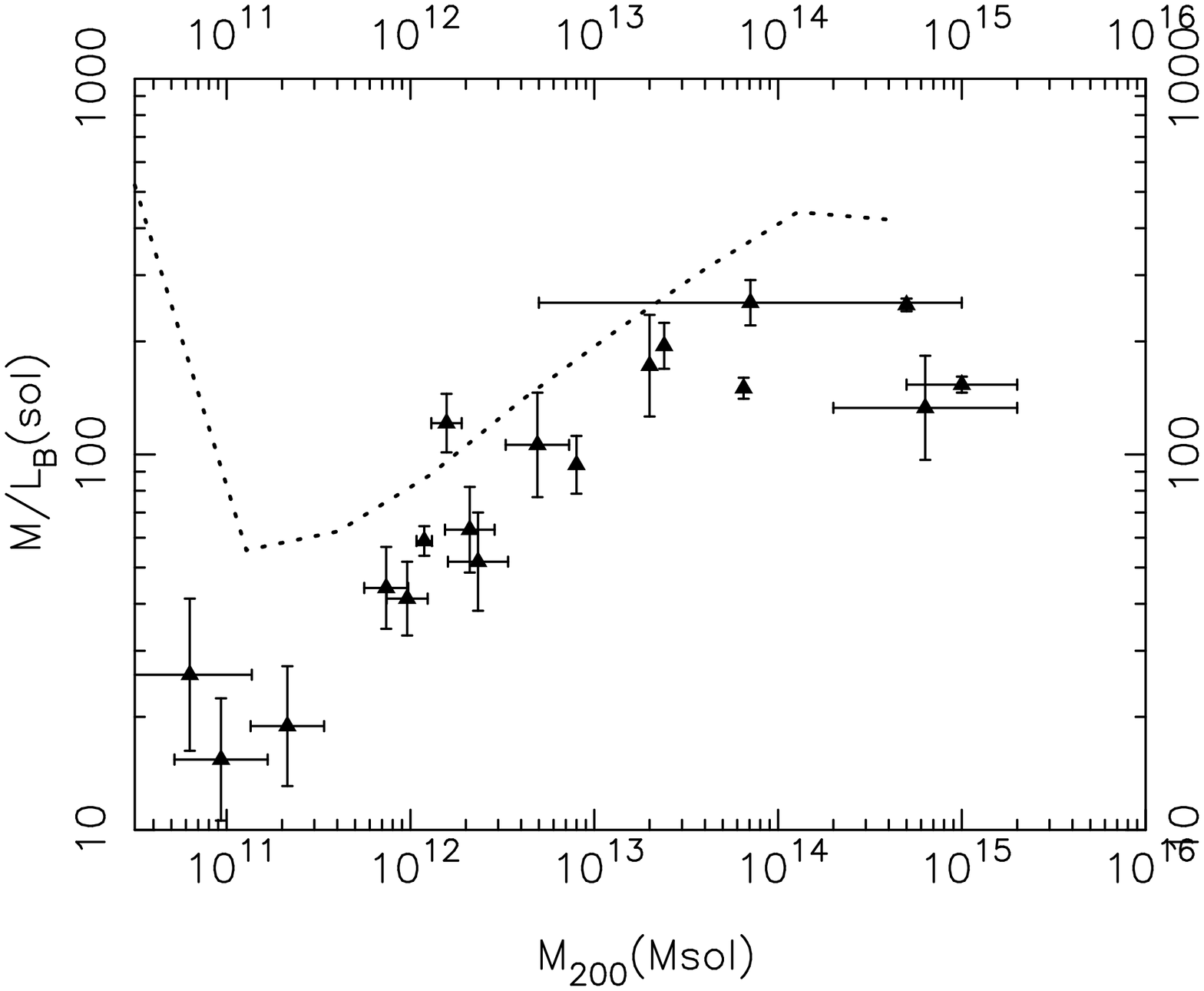}
\end{flushright}
\caption{The amount of total mass in astrophysical bodies.
\textit{Left:} The number of galaxies with a given infra-red $K$-band luminosity in the 2dF 
\cite[circles]{cole01}, SDSS \cite[squares]{kochanek01} and local $z<0.1$ surveys \cite[stars]{huang02}. 
In contrast to this, the dashed line shows a theoretical model of the number of dark matter haloes as a 
function of mass, assuming only cold dark matter physics in the growth of structure \cite{benson03} (see also 
\cite{marioni02}).
This has been converted into luminosity assuming a fixed mass-to-light ratio. 
Its normalisation is arbitrarily chosen to match at the knee of the luminosity function, but can be adjusted 
by changing the model mass-to-light ratio. Importantly, the disparity indicates that baryonic physics act 
to suppress star formation in low-mass or high-mass haloes, and that these contain a very large proportion 
of mass that does not shine.
\textit{Right:} Mass-to-light ratio as a function of the mass (all measured within the radius at which the total density is 200 times higher than the mean density in the Universe), from 
\cite{hoekstra05,gavazzi07,parker05,bardeau07,sheldon07,limousin09,limousin09a,dahle04,heymans08,parker07,
hoekstra04}. The majority of the luminosity 
measurements are made in the B band, at redshift $z\sim0.3$ 
The dotted line shows the prediction of semi-analytic models of galaxy formation \cite{vandenbosch03}.
\label{fig:models_123}} 
\end{figure}

Larger structures have grown through the gradual merger of small structures -- which deepened the 
gravitational potential well, and accelerated the accretion of more mass into runaway collapse. 
According to the Sheth-Tormen/elliptical collapse model of structure formation \cite{sheth01}, 
10\% of the total mass at the present day is contained within galaxy clusters over $10^{14}~M_\odot$ and 
another 15\% within galaxy groups down to $10^{12}~M_\odot$\footnote{To include half of the mass, it is necessary to consider haloes of $10^{10}~M_\odot$,
and 20\% of mass has yet to find its way into a bound halo at all. 
This is much less than in the older Press-Schechter/spherical collapse model \cite{press74}, 
in which 50\% of mass was thought to be in groups and clusters.}. 
This non-linear density enhancement exaggerated the dynamic range of mass fluctuations from the early universe, 
which began with a Gaussian distribution to a high level of accuracy. 
The most massive clusters today are very rare and, since only slightly less dense initial fluctuations grew more slowly,
the present number of haloes of a given mass forms a 
steep ``mass function'' $N(M)$, shown in figure~\ref{fig:models_123}. 
This steepness means that the growth of clusters over time, $N(M,z)$, is very sensitive to the collapse process,
including the nature of gravity \cite{thomas08,guzik09} as well as the amount and physics of dark matter \cite{schuecker03}. 
Conveniently, the dense concentrations of mass also create the strongest gravitational lensing signal.

\begin{figure}[!t]
\begin{flushright}
\includegraphics[width = 197 pt]{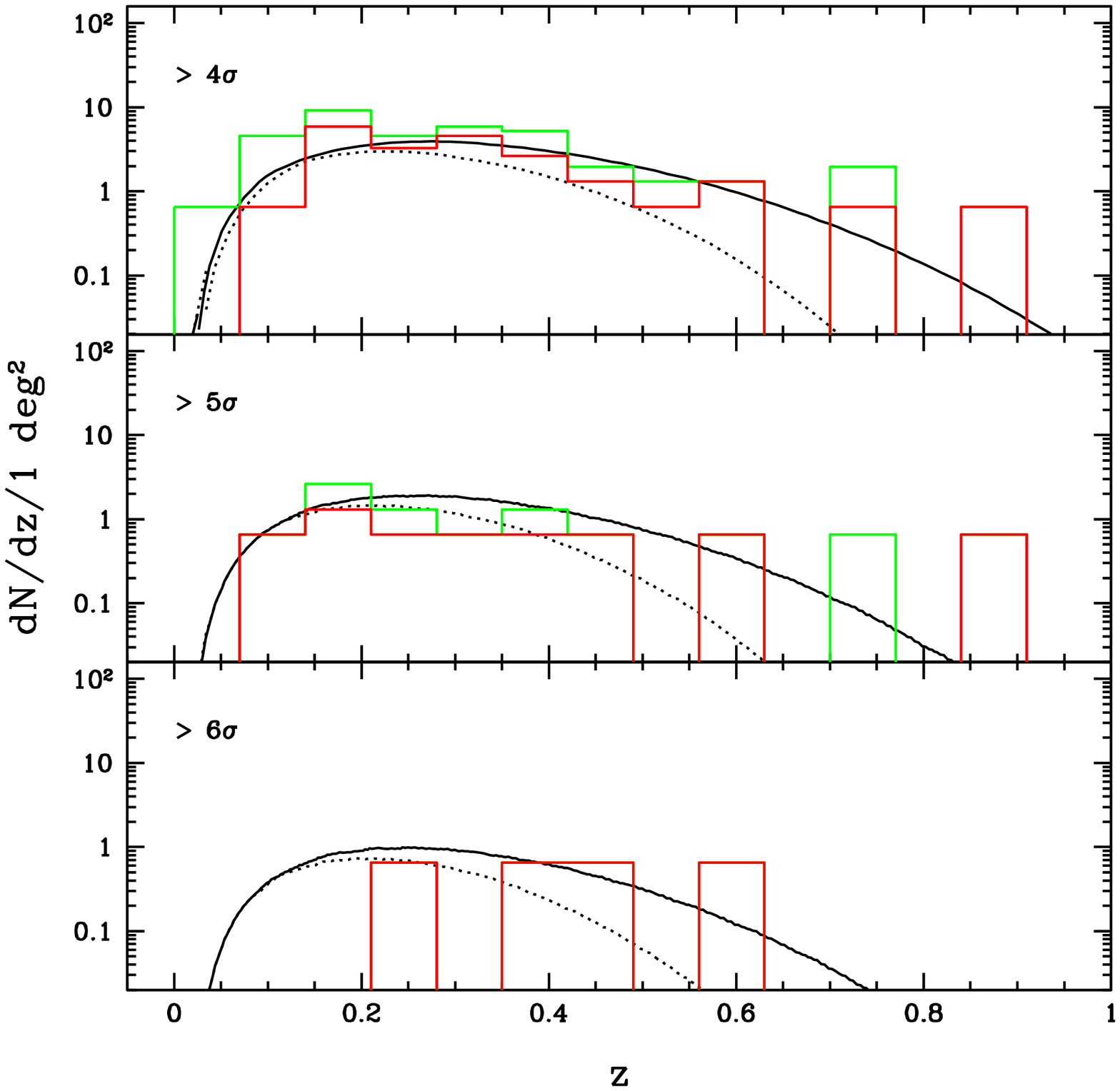}
\includegraphics[width = 192 pt]{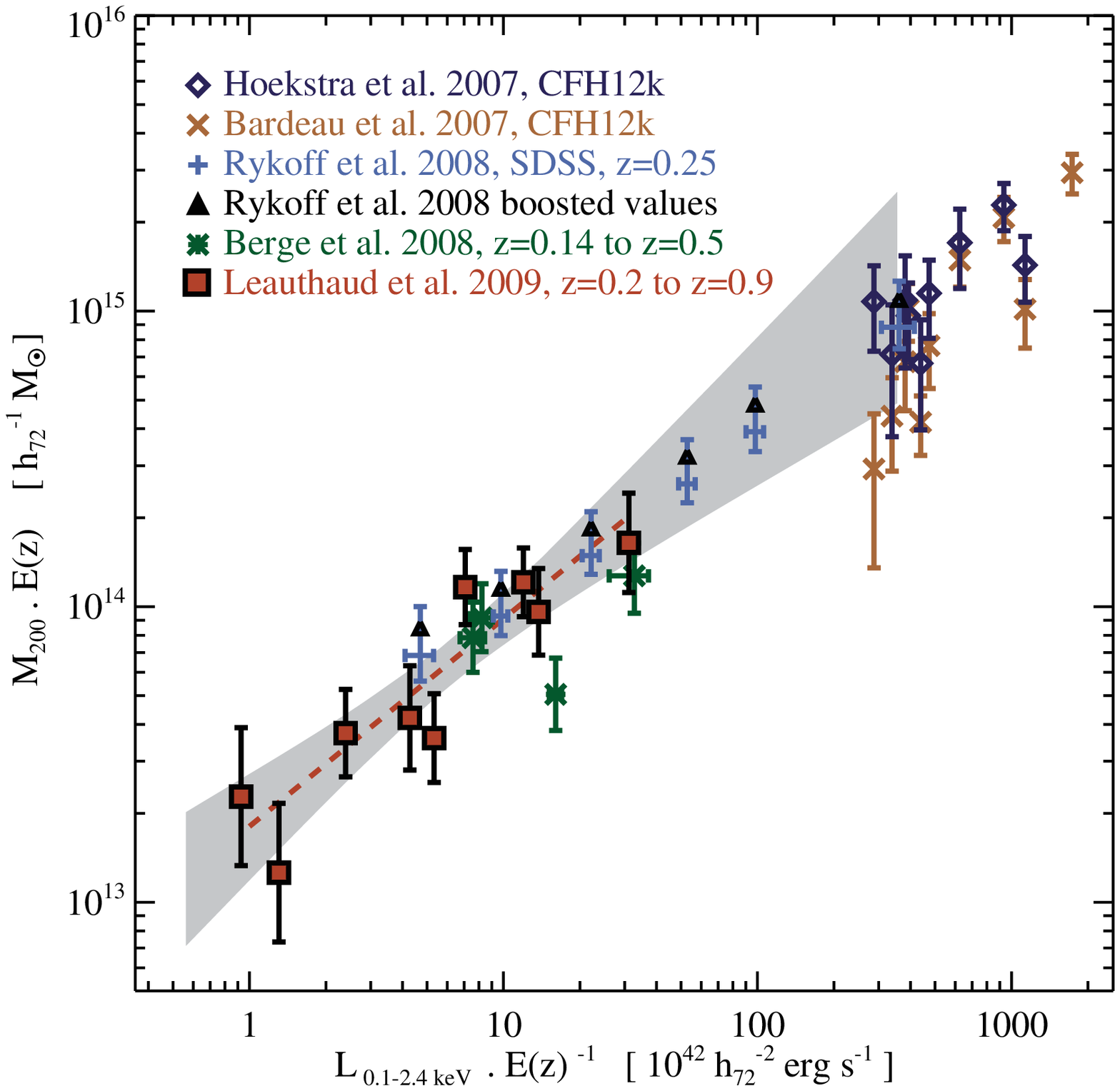}
\end{flushright}
\caption{Counting the number of clusters in the universe $N(M,z)$, as a function of their redshift $z$ and mass $M$. 
\textit{Left:} Directly detecting clusters via their weak gravitational lensing signal, which probably provides the
cleanest selection criteria, using the Subaru telescope \cite{taylor09}. The three panels show different cuts in
detection S/N, which is a proxy for mass $M$. The red histogram shows clusters with spectroscopically confirmed redshifts, the green histogram
shows less secure clusters detected in weak lensing but not yet confirmed, and 
the solid line shows the expected distribution. 
The matching of weak lensing peaks with a baryonic counterpart
requires a large investment of follow-up telescope time, and is the current limitation to the method. There are
currently fewer confirmed clusters than expected, and there is considerable shot noise. However, this technique
shows great promise for the future, with dedicated wide-field surveys.
\textit{Right:} Using weak lensing measurements of a subset of galaxies, groups and clusters to calibrate other
observables -- in this case the X-ray luminosity -- which can then be used to estimate $N(M,z)$ more cheaply.}\label{fig:nmz}
\end{figure}

Galaxy groups and clusters can be found directly via gravitational lensing surveys \cite{faure08,marshall09}. 
Clusters sufficiently massive to produce
strong lensing are generally already known because of the corresponding overdensity of galaxies, although the detection criteria for lensing
is a cleaner function of mass. Weak lensing cluster surveys are advancing even more rapidly.
Several hundred cluster candidates have now been found in weak lensing mass maps from the Canada-France-Hawaii
telescope \cite{gavazzi07b} and the Subaru telescope \cite{miyazaki07}. Follow-up spectroscopy \cite{hamana09} has identified the 
baryonic component of around $60\%$ of these, yielding the redshifts required to place the clusters in $N(M,z)$ plane shown in 
figure~\ref{fig:nmz}, calibrate their mass 
through the geometrical distance to the background galaxies, and also to rule out false detections due to the chance alignment of multiple small
structures along one line of sight \cite{hamana02}.
The remaining $\sim40\%$ of candidates are possibly chance alignments of unrelated small structures, 
or the random orientation of aspherical haloes along the line of sight. Such effects must be
carefully considered in lensing surveys, which are sensitive to the total integrated mass along a line of sight \cite{corless1,corless2}.
Multicolour imaging is also needed to properly identify a clean sample of source galaxies {\it behind} the cluster. Galaxies
inside or in front of the cluster are not lensed by it, and a study of the nearby Coma cluster \cite{gavazzi09} also shows that
member galaxies may even be radially aligned within it, so they will dilute the signal if they are 
misidentified and accidentally included \cite{medezinski09}. 



Gravitational lensing cluster surveys are clean but costly, since it is necessary to find and resolve galaxies more distant than the
structures of interest. 
The baryonic components of clusters can be quickly identified
from infra-red emission, which traces old stellar populations and is unobscured by dust \cite{lin06},
the X-ray luminosity and temperature of intra-cluster gas \cite{vikhlinin06}, the Sunayev-Zeldovich (SZ) effect in which the CMB is
scattered to higher energy off warm electrons \cite{bn1,bn2}, and Doppler-shifted light that
reveals the clusters' internal kinematics \cite{blindert04}. 
In particular, since X-ray emission is proportional to the square of the electron density in intra-cluster gas,
X-ray surveys are less sensitive than lensing to the chance alignment of many small haloes along a line of sight.
However, the fundamental quantity most easily predicted by theories
is mass, and a scaling relation must be constructed from all of these luminous observables to mass.
The scaling relations often rely on poorly justified assumptions about the dynamical equilibrium
or physical state of the baryonic component \cite{fort94,allen98}.
Inherent systematic errors can be investigated by an inter-comparison of the various observables, 
but the ultimate comparison is now generally obtained versus gravitational lensing \cite{bardeau07,hoekstra,berge08,marrone09,mcinnes09}. 
Strong lensing arcs directly measure the
enclosed mass within the Einstein radius, providing a robust normalisation of the mass
distribution, and weak lensing traces the outer profile of the halo, where most of the mass is found.

The comparison with X-ray cluster surveys has been most astrophysically interesting.
The combination of strong lensing and X-ray measurements of galaxy clusters was first advocated as a way
to probe the dynamics of the intra-cluster gas \cite{meab}.
Initial disagreements in the overall normalisation \cite{smail98,allen01} have indeed been much addressed by
accounting for the effects of cool cluster cores on emission from the intra-cluster medium. 
Finally, a comparison was completed of strong lensing, X-ray and infra-red emission from  
$10$ X-ray luminous (L$_{\rm X}$ $>8\times10^{44}$~ergs/s at $0.2$-$2.4$~keV inside $R<350$~kpc) clusters at redshift
$z\sim0.2$ \cite{smith05}. 
As shown in figure~\ref{fig:nmz}, mass measurements now generally agree for dynamically mature clusters 
with a circular X-ray morphology and high central concentration of the infra-red light.
However, at a certain level, there is no such thing as a relaxed cluster.
Major mergers leave more than half of systems dynamically immature, and estimates of their mass from the complex X-ray morphologies
are particularly problematic \cite{bower01}. In these cases, only lensing mass estimates appear viable.

\subsection{Amount of dark matter in large-scale structure}

\begin{figure}[!t]
\begin{flushright}
\includegraphics[scale = 0.37]{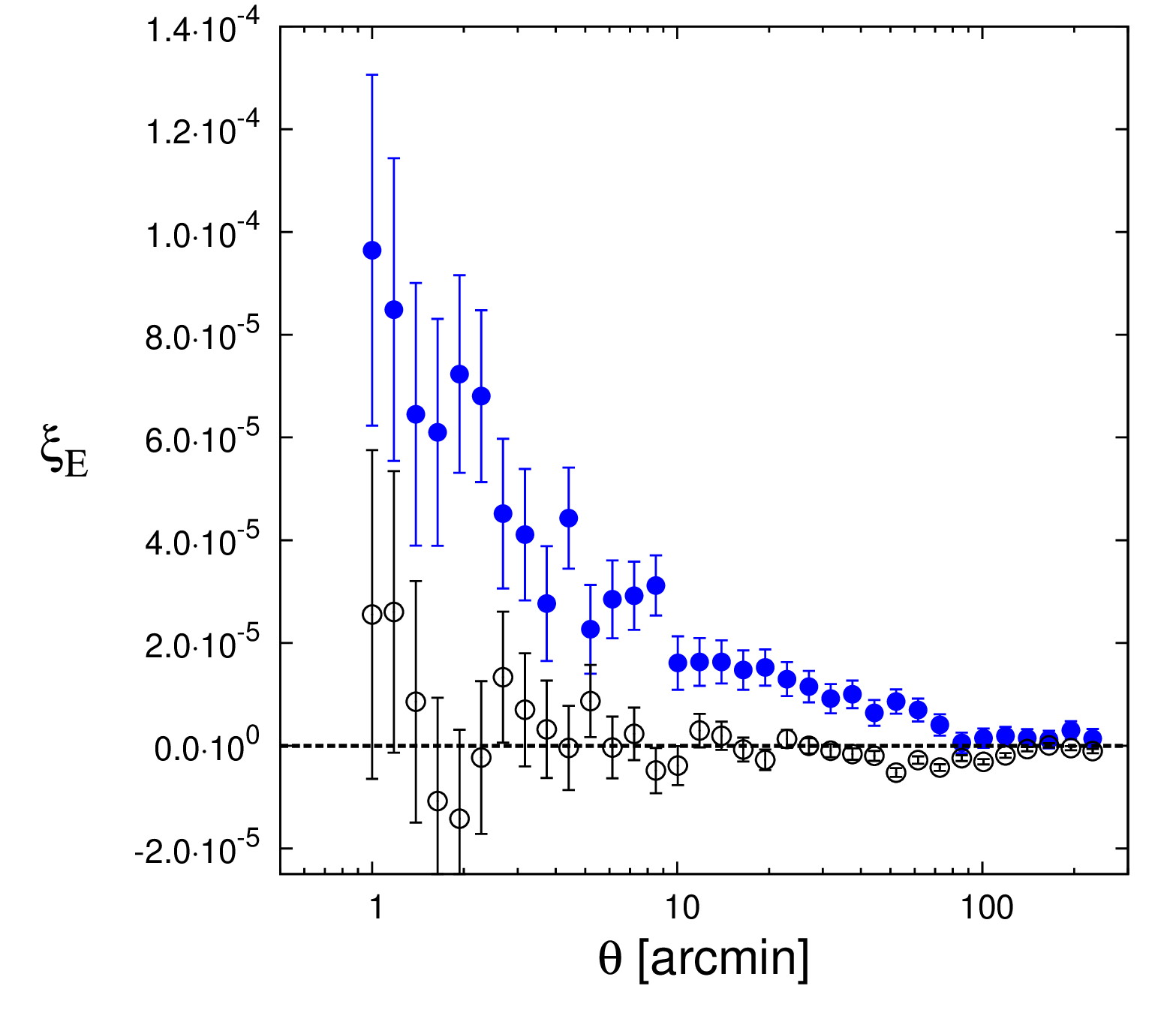}
\includegraphics[scale = 0.26]{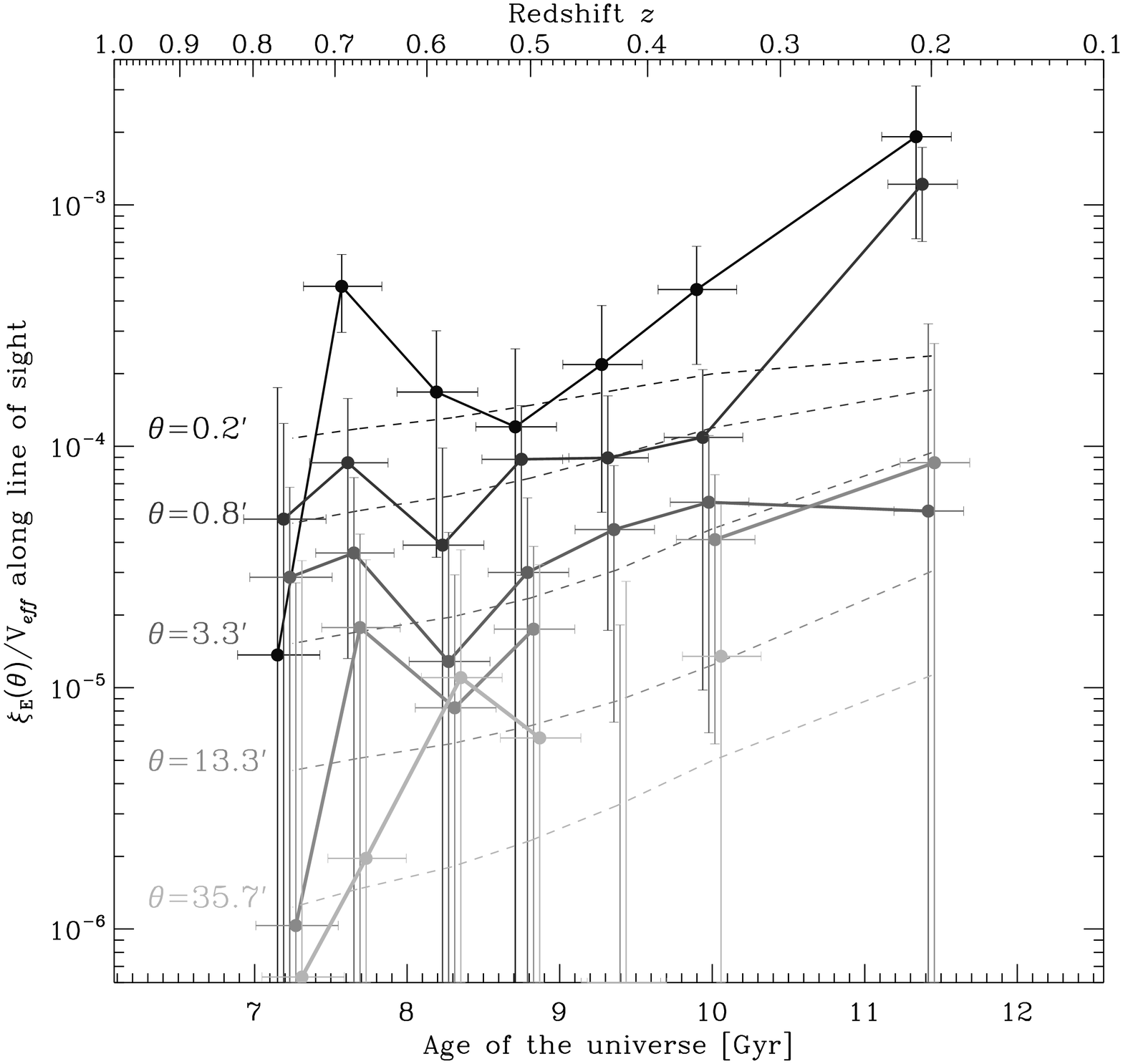}
\end{flushright}
\caption{The large-scale weak lensing ``cosmic shear'' signal.
\textit{Left:} Measurement of the 2D signal from the Canada-France-Hawaii telescope Legacy
Survey \cite{fu08}. This traces the overall amount of mass in the Universe, projected along the line of sight, 
and shows how
it is more clumped on small scales than on large scales. Solid points show the cosmological $E$-mode lensing signal, and
open points show the $B$-mode, a tracer of uncorrected systematic effects that should be consistent with zero. 
\textit{Right:} The growth of
this signal over cosmic time, measured from the Hubble Space Telescope COSMOS survey \cite{massey07b}. This
uses the 3D locations of source galaxies to trace the distribution of mass at different distance from the
Earth. Dashed lines show the prediction of the standard $\Lambda$CDM cosmological model. Error bars account
for only statistical error within the field and do not include the effect of using only a small field.}\label{fig:cosmicshear}
\end{figure}

\begin{figure}[!t]
\begin{flushright}
\includegraphics[width = 190 pt]{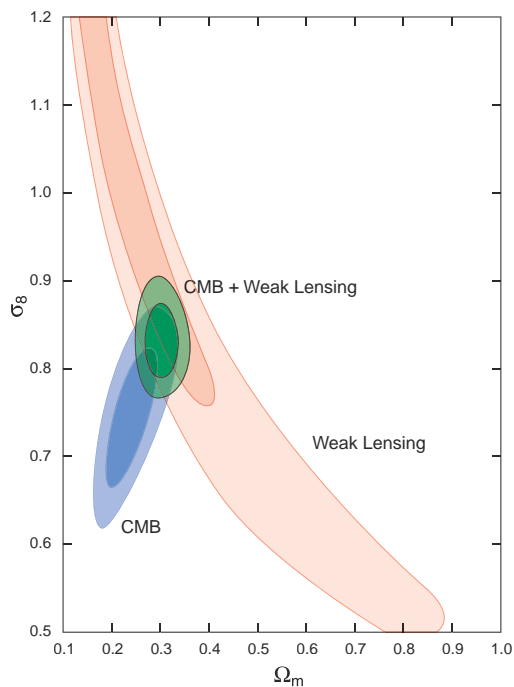}~~~~~~~~~~~~~~~~~~~~~~~~~
\end{flushright}
\caption{Constraints on cosmological parameters from the distribution of
mass in the early universe, traced by the WMAP satellite in the Cosmic Microwave Background radiation, compared 
to the distribution of mass in the local universe from weak gravitational lensing \cite{spergel07}. The parameter $\Omega_m$ is the density of all mass in
the Universe, and $\sigma_8$ is the normalisation of the power spectrum, describing its clumpiness on $8h^{-1}$~Mpc
scales. In each case, the two contours depict $68$\% and $95$\% confidence limits, and assume a flat universe.
The orthogonality of the two constraints, originating from the evolution of the mass distribution between very 
different epochs, is key to their combined power.}\label{fig:wmap}
\end{figure}

Large weak lensing surveys of ``cosmic shear'' along random lines of sight
can be used to study the distribution of mass on the largest scales, and the mean density of the Universe 
($\Omega_m$, which is usually expressed in units of the fraction of the density required to {\it just} close the 
Universe and prevent perpetual expansion).
The amount of mass clumped on different scales is usually parameterised in terms of
the (two-point) correlation $\xi_E$ between the cumulative shear distortion along lines of sight to
pairs of galaxies separated by an angle $\theta$ on the sky, as illustrated in figure~\ref{fig:cosmicshear}. 
In isolation, current cosmic shear constraints on $\Omega_m$ are degenerate 
with $\sigma_8$ (see figure~\ref{fig:wmap}), another parameter in cosmological models that normalises the amount of clumping of matter 
on a fixed scale of $8h^{-1}$~Mpc --  in this sense, it describes the physical size of the clumps.
At a redshift $z=0.3$, where many recent cosmic shear surveys are most
sensitive, $8h^{-1}$~Mpc corresponds to an angular size of $\sim43$~arcminutes on the sky (about one and a half times
the diameter of the full moon). The degeneracy between
$\Omega_m$ and $\sigma_8$ is gradually being removed, as larger cosmic shear surveys 
measure probe the distribution of dark matter with statistical significance on both larger and smaller scales. 
Extensions towards large scales are particularly welcomed, because very large-scale structure is still collapsing linearly, so
theoretical predictions are calculable from
first-order perturbation theory. The degeneracy is also being broken by the first measurements of the three-point
correlation function of galaxy triplets \cite{bardeau07,semboloni09,berge09}, which is sensitive to the skewness
of the mass distribution, and depends in an orthogonal way upon $\Omega_m$.

The ``clumpiness'' of matter naturally increases as the Universe transitions
from an almost uniform state at high redshift to the structures we see around us today. 
The rate of growth of this structure also depends upon $\Omega_m$, since additional mass speeds up gravitational collapse.
The degeneracy between $\Omega_m$ and $\sigma_8$ present in a static, 2D analysis can this be broken by comparing
the density fluctuations at different epochs.
Figure~\ref{fig:cosmicshear} shows contraints of $\Omega_m=0.248\pm0.019$  
from a comparison of the primordial matter fluctuations captured in the Cosmic Microwave
Background radiation \cite{spergel07} with current structure seen in
weak lensing measurements from the $50$ square degree patch Canada-France-Hawaii telescope Legacy Survey \cite{fu08}. 

Even tighter constraints on $\Omega_m$, and unique insight into the nature of gravity as it shapes dark matter,
can be obtained by tracing the continual growth of structure \cite{bacon04}. 
This can be obtained from gravitational lensing because, while nearby galaxies are lensed by local structure between them and us, more
distant galaxies are also lensed by the additional mass in front of them, and the most distant galaxies are
lensed by mass throughout the Universe. The finite speed of light makes distance equivalent to
lookback time, so we can reconstruct the distribution of mass in distant structures as it was when the light passed near
and was lensed by them many billions of years ago. 
Redshifts can be used as a proxy for the distance to each lensed galaxy, and are measured from the spectrum of their emitted light or estimated from multicolour images. 
The $2$ square degree Hubble Space Telescope COSMOS survey is the largest optical survey ever obtained from space, 
with extremely high quality imaging that resolves the shapes of even small and faint galaxies at lookback times of more than 10 billion years.
Multiwavelength follow up of the field in about forty other wavelengths, from radio, through IR, optical, UV and X-ray, 
provides the most accurate photometric redshift estimates, for about 2~million galaxies \cite{capak07}.
The right hand panel of figure~\ref{fig:cosmicshear} shows the essentially independent measurements of $\xi_E(\theta)$ as a function of time, witnessing the growth of structure.
Compared to a 2D analysis, this tightens statistical errors on $\Omega_m$ by a factor of $3$ \cite{massey07b}, yielding
$\Omega_m=0.247\pm0.016$ from only a $2$ square degree patch of sky \cite{lesgourgues07}. 
A continuous 3D cosmic
shear analysis can potentially provide five-fold improvements over a 2D survey \cite{heavens03,kitching07}, making
the investment of follow-up telescope time very effective.

Galaxies can only be resolved to finite distances, and they did not even exist in the very early Universe. 
As well as providing a snapshot of primordial density fluctuations, the CMB may also provide the
ultimate high redshift source that has been gravitationally lensed by even more 
foreground matter \cite{hu01,lesgourgues06,lewis06}. 
Patterns in the temperature of the CMB form shapes that become distorted by lensing in exactly the same way 
as galaxies. More interestingly, lensing moves CMB photons without rotating their polarisation. 
Primordial density fluctuations from a scalar inflationary field produce a curl-free ($E$-mode)
polarisation signal, but this is mixed by lensing into a non-zero $B$-mode signal \cite{efstathiou09}.
This requires high angular resolution measurements of the CMB and 
only upper limits have yet been measured \cite{quad}. 
To complicate matters further, non-zero curl modes in the polarisation can also be created by
foreground effects such as dust emission, and tensor (gravity-wave) perturbations in the early Universe.


Most ambitiously, independent measurements of $\Omega_m$ at vastly different cosmic epochs
could also constrain the conservation of mass in the Universe, although the statistical error on this
are likely to remain large for the foreseeable future.


\section{Organisation of dark matter} \label{sec:organisation}

\subsection{Distribution on large scales}

\begin{figure}
\begin{flushright}
\includegraphics[width = 180 pt]{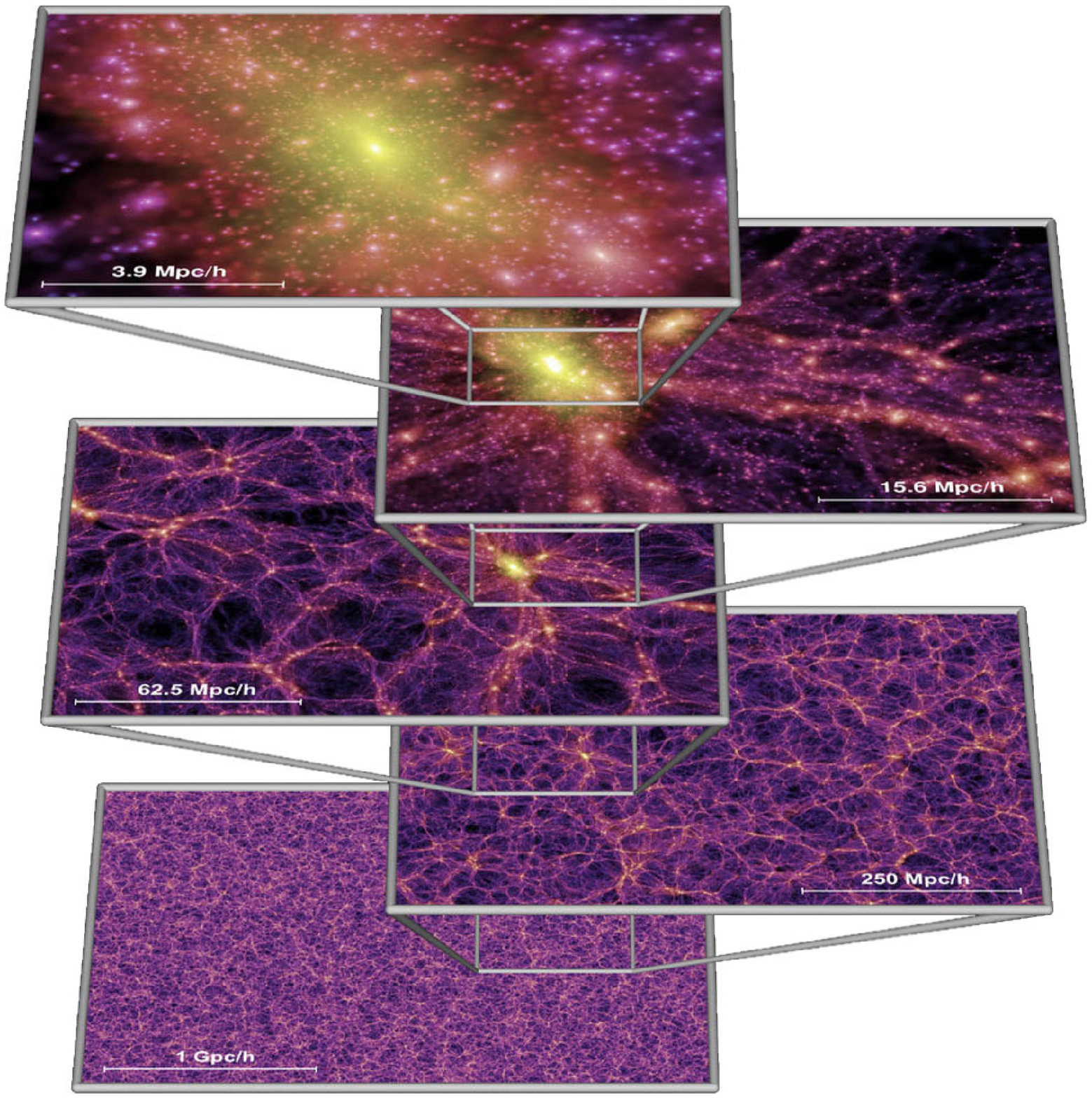}~~
\includegraphics[scale = 0.32]{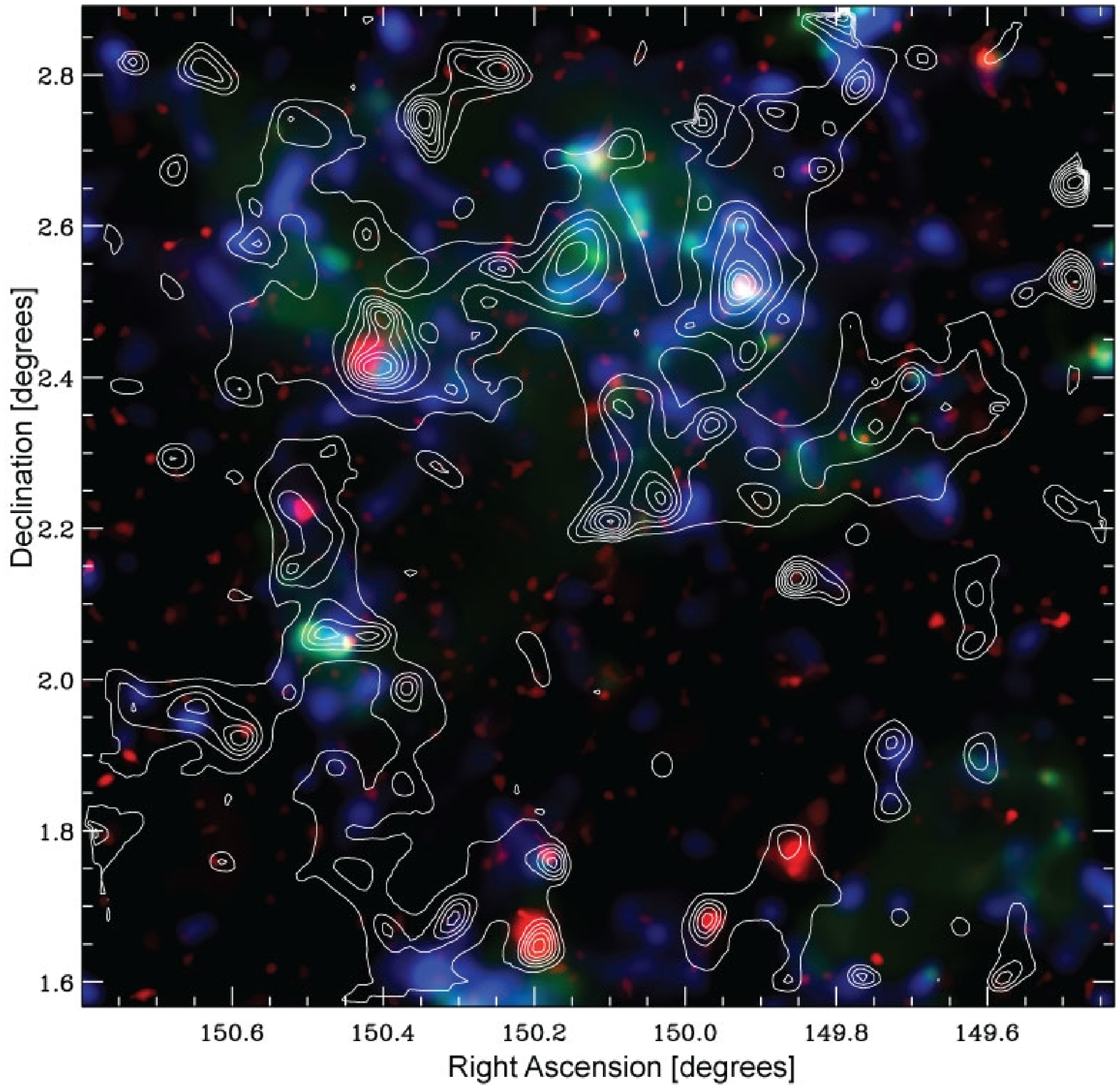}
\end{flushright}
\caption{The large-scale distribution of dark matter.
{\it (Left):} The expected distribution of dark matter at the present day, from the ``Millenium simulation'' (of
a Universe containing only cold dark matter) \cite{springel05}. Each layer is zoomed from the last by a factor
of four, and shows the projected distribution of dark matter in slices of $\sim20$~Mpc thickness, colour-coded
by density and local dark matter velocity dispersion (Image credit: Volker Springel/Max Planck Institute for
Astrophysics).
{\it (Right):} The observed large-scale structure in the Hubble Space Telescope COSMOS survey
\cite{massey07a}. Contours show the reconstructed mass from weak gravitational lensing, 
obtained by running the filters in the left hand panel of
figure~\ref{fig:eb} across the observed distribution of background galaxies shown the right hand panel of figure~\ref{fig:eb}
(to improve resolution, the conversion process actually used smaller, noisier bins each containing $\sim80$
galaxies). Like an ordinary optical lens, a gravitational lens is most effective half-way between
the source and the observer. At redshift $z\sim0.7$, where the lensing measurements are most sensitive, the
field of view is about the same as the second layer from the top in the left panel. However, the observations also include
overlaid contributions (at a lower weight) from all mass between redshifts $0.3$--$1.0$ along our line of sight, projected
onto the plane of the sky.  The various background colours depict different tracers of baryonic mass.  Green
shows the density of optically-selected galaxies and blue shows those galaxies, weighted by their stellar mass
from fits to  their spectral energy distributions. These have both been weighted by the same sensitivity
function in redshift as that inherent in the lensing analysis. Red shows X-ray emission from hot gas in
extended sources, with most point sources removed. This has not been rewighted, so is stronger from nearby sources, and weaker from the
more distant ones.}\label{fig:cosmosmap}
\end{figure}

On the largest scales, dark matter forms a crisscrossing network of filaments, spanning vast, empty voids, and
with the largest concentrations of mass at their intersections. Figure~\ref{fig:cosmosmap} compares the
expected and observed distribution of mass. The $2$ square degree Hubble Space Telescope COSMOS survey was 
specifically designed to enclose a contiguous volume of the universe at
redshift $z=1$ containing at least one example of even the largest expected structures \cite{scoville07}.
The filamentary network of mass revealed by weak lensing measurements is apparent, and the multiwavelength imaging 
also provides several tracers of baryons \cite{capak07}. Optical and infra-red
emission, when interpreted via theoretical models of stellar evolution, can be used to infer the mass, age
and other properties of the star populations. X-ray imaging is sensitive to the gas in dense clusters of
galaxies that is heated sufficiently for it to glow at these shorter wavelengths. 

The multicolour data also provide redshift estimates for each source galaxy that can be used to extrude the observed
map in 3D along the line of sight \cite{bacon03,massey07a}. This technique has also been applied to the distribution
of dark matter near the Abell~901/902 galaxy supercluster \cite{simon09}, resulting in the discovery of a previously 
unknown cluster CB1 that lay behind the foreground \cite{taylora04}.

\begin{figure}[!t]
\begin{flushright}
\includegraphics[scale = 0.34]{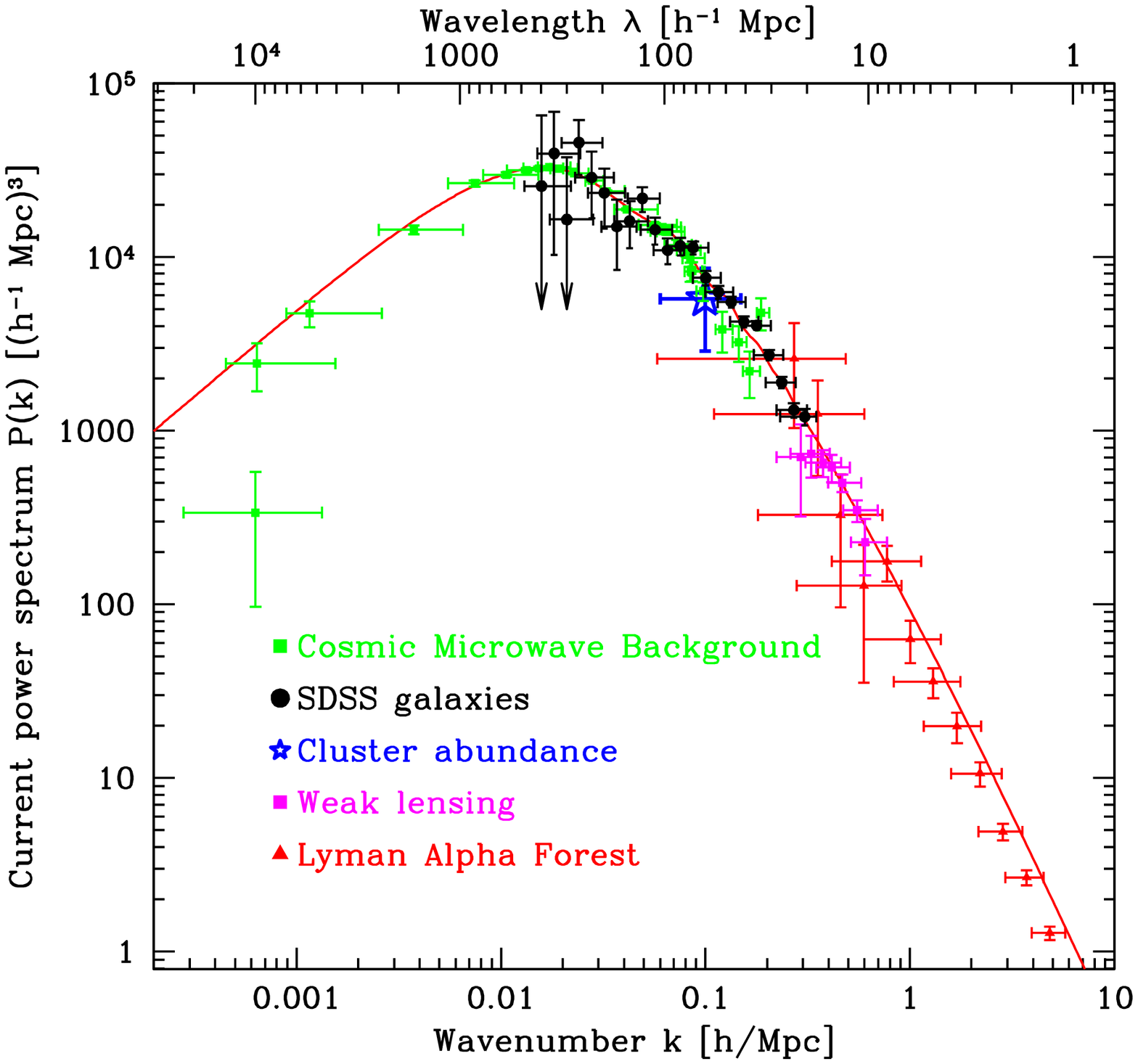}
\includegraphics[scale = 0.33]{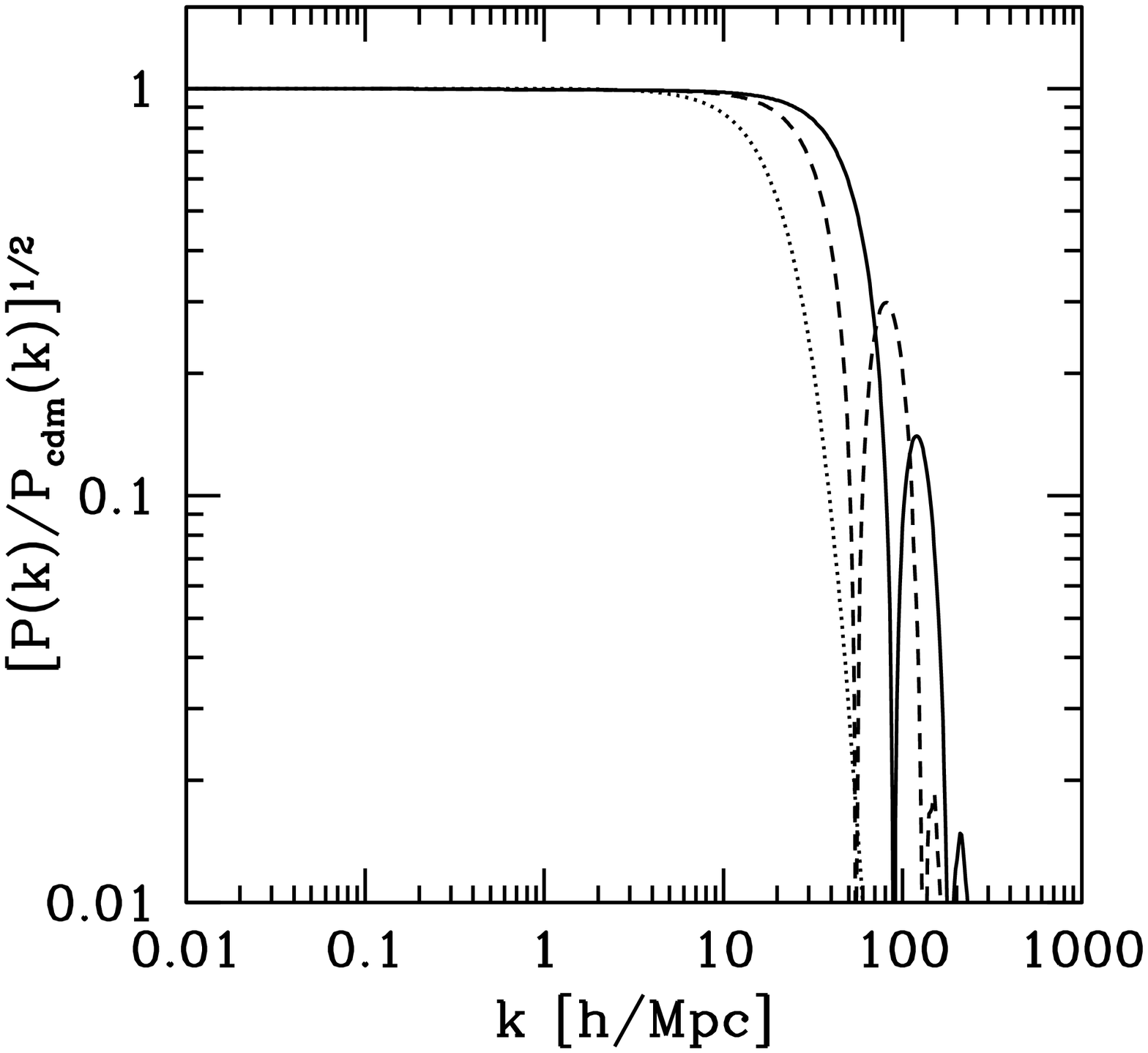}
\end{flushright}
\caption{The statistical distribution of dark matter in Fourier space.
\textit{Left:} The mass power spectrum, showing the clumping of dark matter as a function of 
large scales (low $k$) to small scales (high $k$) \cite{tegmark04}, including some very early weak lensing constraints \cite{hoekstra02}.
Current constraints are tighter and extend over a wider range of scales in both directions.
\textit{Right:} The effect on the power spectrum of 1~MeV mass WIMP dark matter that remains coupled to other particles in the
early universe \cite{hooper07}. The solid (dashed) line assumes a  10~keV (1~keV) kinetic decoupling temperature.
The dotted line illustrates the similar small-scale damping effect of a component of warm dark matter.}
\label{fig:powerspect}
\end{figure}

The large-scale distribution of mass can be described statistically in terms of its power spectrum $P(k)$, which
is shown in the left hand panel of figure~\ref{fig:powerspect}. This is the Fourier transform of the correlation functions shown in
figure~\ref{fig:cosmicshear}, and measures the amount of clumping on different physical scales. If the density
field is Gaussian, the same information can also be expressed as the mass function $N(M,z)$, i.e.\ the density
of haloes of a given mass as a function of that mass and cosmological redshift. There has been a substantial
amount of work on analysing the properties of the dark matter power spectrum, including its growth over time from
a power spectrum of primordial density fluctuations.
In order to compare theoretical
models to data, the semi-analytic approach of \cite{Eisenstein} is used, predominantly for its simple
fitting functions to the linear power spectrum, which can be extended (at $5$-$10$\% accuracy) into the mildly 
nonlinear high-$k$ regime \cite{pandd,smith}. 

In the early universe, photons were subject to density waves in which regions could gravitationally collapse
then, with the increased density providing pressure support, rebound and oscillate. Large regions oscillated
slowly, but smaller regions could complete multiple cycles. The oscillations were frozen when the temperature of
the Universe dropped sufficiently for protons to capture electrons, and fluctuations that happened to be
particularly overdense or underdense are now seen as anisotropies in the Cosmic Microwave Background. Since
standard model particles were coupled to photons at high energies, they were subject to the same density
fluctuations, and also froze out at decoupling. To this day, baryonic structures like galaxies exist preferentially
at certain fixed physical separations, known in the power spectrum as Baryon Acoustic Oscillations (BAOs). Dark matter
was initially not subject to these fluctuations, having a featureless power spectrum. Indeed, it
decoupled from the primeval soup (of photons and particles) at very early times, 
and formed the first network of structures that acted as
scaffolding into which baryonic material could be drawn and assembled, then only later picking up the preferred scales 
through gravitational interaction with the baryons. However, in several candidate particle 
models, including supersymmetric particles, dark matter couples to photons at energies even 
higher than the standard model particles. The smallest
density fluctuations would have had chance to oscillate once or twice, and imprint their scales on even the
primordial distribution of dark matter, as shown in the right hand panel of figure~\ref{fig:powerspect} \cite{hooper07}. In addition, a warm or hot component dark matter, such as
low-mass neutrinos, could continuously free-stream away from any mass concentrations that build up (refusing to be captured in
dense regions of the Universe) and erase structure on small scales (see Section \ref{sec:neutrino} for more details).

\subsection{Sizes of individual haloes}


In numerical simulations, collisionless dark matter particles form  structures with a remarkable, ``universal''
density profile across a wide range of mass scales from dwarf galaxies to clusters. There may be some 
cluster-to-cluster scatter \cite{jing00,madau08,diemand08}, but the mean density profile $\rho(r)$ is expected to rise as
$\rho\propto r^{-3}$ in the outskirts, transitioning to $\rho\propto r^{-\beta}$ inside a scale radius $r_s$ that
is a function of halo mass and formation redshift. Early simulations found a central cusp with $\beta$ 
between $1$ and $1.5$ \cite{nfw,moore99}, but this appears to have been an effect of their limited resolution, with more recent
simulations predicting a smooth decrease in slope towards a flat core \cite{navarro08}.
This appears to be converging, but thorough testing of the numerical simulations is continuing.
Finalising and comparing these predictions to the observed distribution of dark matter in real
clusters is a strong test of the whole collisionless CDM paradigm -- as well as the nature of gravity.

\begin{figure}[!t]
\begin{flushright}
\includegraphics[width = 372 pt]{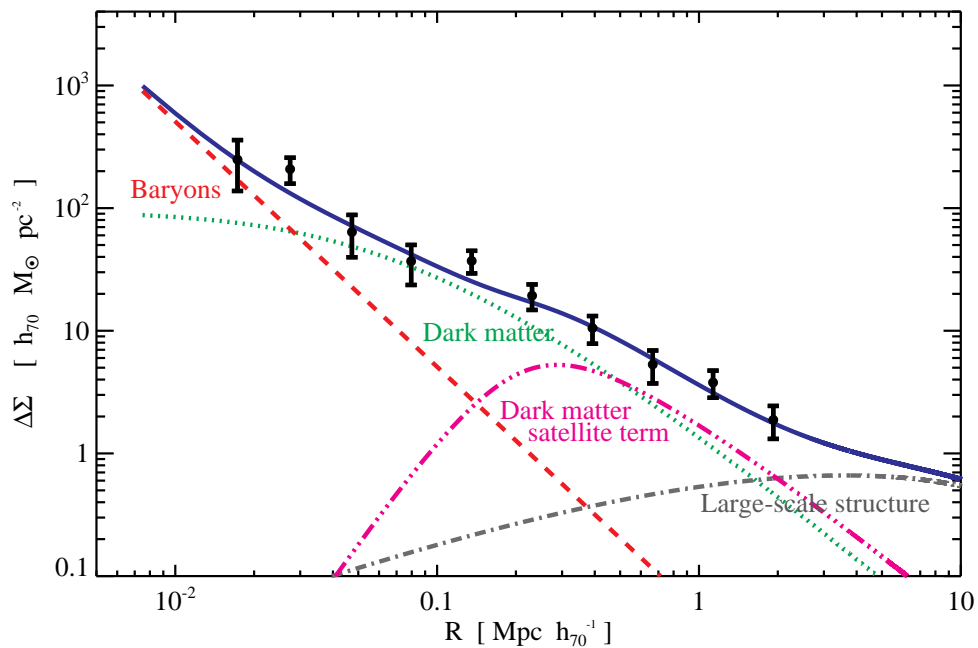}
\end{flushright}
\caption{The observed radial distribution of mass around elliptical galaxies in the Hubble Space Telescope
 COSMOS survey, decomposed into its various components \cite{leauthaud09}. 
 The solid blue curve shows the total ``galaxy-galaxy'' weak gravitational lensing signal. 
 On small scales around $\sim 10$~kpc, this is dominated by the baryonic content of galaxies
 represented by the red dashed curve. This particular data set is for a set of elliptical galaxies whose
 spectral energy distributions indicate a similar amount of mass in stars, and the central lensing signal
 increases as expected for galaxies with larger predicted stellar masses. At intermediate scales
 around $\sim 200$~kpc, dark matter haloes become dominant: the main NFW halo term (dotted green), 
 plus an additional contribution (triple-dot-dash magenta) from occasions when the analysis focuses on 
 satellite galaxies in the halo of a larger host, rather than main galaxies. 
 On large scales above 3~Mpc, the galaxy-galaxy lensing signal reverts to the cosmic shear signal from
 large-scale structure in which the galaxy is located (dot-dash grey).}\label{fig:gg_profile}
\end{figure}

Individual galaxies of $10^{12}M_\odot$ at $z\sim0.22$ have weak lensing signals that show 
extended dark matter haloes, with large scale radii
in agreement with simulations \cite{mandelbaum06b}. Note that the astrophysics literature normally parameterises this
in terms of the `concentration' or ratio of the radius containing most of the mass to the scale radius 
(individual galaxies are expected
to have concentrations of around $7$--$13$, and massive clusters of $5$--$6$). A sample of $98$ galaxies acting as strong
lenses was ingeniously found by looking for multiple sets of spectral lines at different redshifts within sources that 
need not even be resolved in low resolution imaging \cite{bolton04,bolton08}. 
A successful observing campaign with the Hubble Space Telescope has now followed up $\sim150$ such targets, with a $2$ in $3$
success rate of resolving strong lens systems \cite{gavazzi07}. Interestingly, the stacked gravitational lensing
signal behind them, in agreement with dynamical analysis, shows an apparent conspiracy between the dark matter and baryonic components to produce an
overall ``isothermal'' $\rho\propto r^{-2}$ density profile out to very large ($\sim140$~kpc) radii
\cite{nipoti08,gavazzi08} (or even further \cite{sheldon04}). Figure~\ref{fig:gg_profile} demonstrates the conspiracy of central
baryons, the galaxy's own dark matter halo, and the haloes of neighbouring galaxies; none of them is individually
isothermal \cite{mandelbaum06, gavazzi07, johnston07, leauthaud09}. In addition, the location of the transition from the host halo to
large-scale structure marks the typical size of dark matter structures and its occurrence at the scale expected
from simulations itself provides strong support for the CDM paradigm \cite{hoekstra05}.

Galaxy clusters of $10^{14}-10^{15}M_\odot$ in the Sloan Digital Sky Survey also exhibit dark matter 
haloes with scale radii $r_s$ in line with expectations \cite{mandelbaum06}. 
Other surveys find particularly
massive clusters to have smaller scale radii (but the expected outer slope), as would have
happened if they had collapsed earlier, when the Universe was more dense \cite{kneib03,gavazzi03,comerford07,broadhurst08,oguri09}. 
The baryons that accumulate in the
cluster cores (not least in the typically massive central galaxy) add a complication that is not included in the
simulations and difficult to disentangle in observations. Baryons form stars and radiate away energy, falling
further into a deepening gravitational potential well that also drags in and increases the central concentration of
dark matter \cite{blumenthal86,gnedin04}. However, additional baryonic effects would act in the opposite way, and 
the discrepancy may simply arise from selection biases that favour the observation of haloes that
are centrally concentrated or triaxially elongated in 3D and oriented along our line of sight
\cite{sheldon08,mandelbaum09}. Nonetheless, these observations suggest the intriguing possibility of 
non-Gaussian density fluctuations in the early Universe (potentially cosmic strings) that would have
seeded accelerated structure formation \cite{mathis04}.

\subsection{Shapes of haloes}

In the standard cosmological model, dark matter haloes are expected to be significantly non-spherical \cite{allgood,jing}.
Measurements of weak gravitational lensing in the Sloan Digital Sky Survey confirm that the axis ratio of haloes
around isolated galaxy clusters (projected onto the $2$D plane of the sky) is $0.48^{+0.14}_{-0.19}$ \cite{evans}.
This rules out sphericity at $99.6\%$ confidence and is consistent with the ellipticity of the cluster galaxy
distribution. Albeit with large statistical errors, the dark matter haloes around individual galaxies appear slightly rounder than the light
emission. In the Canada-France-Hawaii Telescope's (CFHT) Red Cluster Sequence (RCS) survey, 
the mean ellipticity of dark matter haloes around all galaxies is
$77^{+18}_{-21}\%$ that of the host galaxy ellipticity \cite{hoekstra04}. This has been subdivided in SDSS to
haloes around red (elliptical) galaxies with $60\pm38\%$ of the ellipticity of their host galaxies, versus haloes
around blue (spiral) galaxies being \textit{anti-}aligned with their host galaxies and \textit{more} oblate by
$40^{+70}_{-100}\%$ \cite{mandelbaum06a}. This dichotomy is also seen in the CFHTLS \cite{parker07}. Such
results confirm that the dark matter haloes guided the formation of the cores of massive galaxies
\cite{koopmans06}. Constraints on the ellipticity of dark matter haloes may be two orders of magnitude tighter from gravitational
flexion than those from shear \cite{hawken}. As survey get larger and observations improve, the next step will be to test whether the haloes
align with the large-scale structure in which they formed. For example, tidal gravitational forces along filaments
may preferentially align  haloes, elliptical galaxies, and the angular momenta of spiral galaxies
\cite{schneider09}.
It has also been suggested \cite{ho,johnston05} that the average ellipticity of dark matter haloes can be used
to probe cosmological parameter $\sigma_8$ (and to a lesser extent $\Omega_m$), and the nature of gravity.

Very oblate mass distributions
can produce three images of a strongly-lensed source, as opposed to the
usual double or quadruple images. The paucity of observed triple image systems therefore suggests that most are quite
spherical \cite{mandelbaum09} -- although the expected abundance also depends upon the inner profile of the mass
distribution. For strong lenses with a small separation ($\ls5''$), the radial profile and hence the image multiplicities depend
sensitively on the alignment of the dark matter and central baryonic components \cite{minor09}. Conveniently however, 
although the orientation of an individual galaxy (particularly spiral galaxies) with respect to the line
of sight affects its strong lensing cross-section, when averaged over all possible orientations or relative
orientations of dark matter and luminous components, the shape of a spiral or elliptical galaxy does not bias the
strong lensing signal by more than $\sim10\%$ \cite{minor09,keeton98,mandelbaum09}.


Numerical simulations suggest that the dark matter haloes around spiral disc galaxies have an additional
interesting feature. Merging subhaloes approaching from within or near the plane of the disc are
gravitationally dragged into a ``dark disc'' \cite{read08}, which maintains a similar velocity dispersion to
the stars that form the thick disc. A dark disc would have important implications for dark matter direct 
detection experiments, because of the low velocity of those dark matter particles with respect to the Earth.
However, the stellar disc is a small fraction of stellar mass, and it will be challenging
to detect an equivalent dark component via lensing (or dynamical) measurements.

\subsection{Cusp versus core central profiles}\label{sec:cuspcore}

\begin{figure}[!t]
\begin{flushright}
\includegraphics[width = 160 pt, height=160 pt]{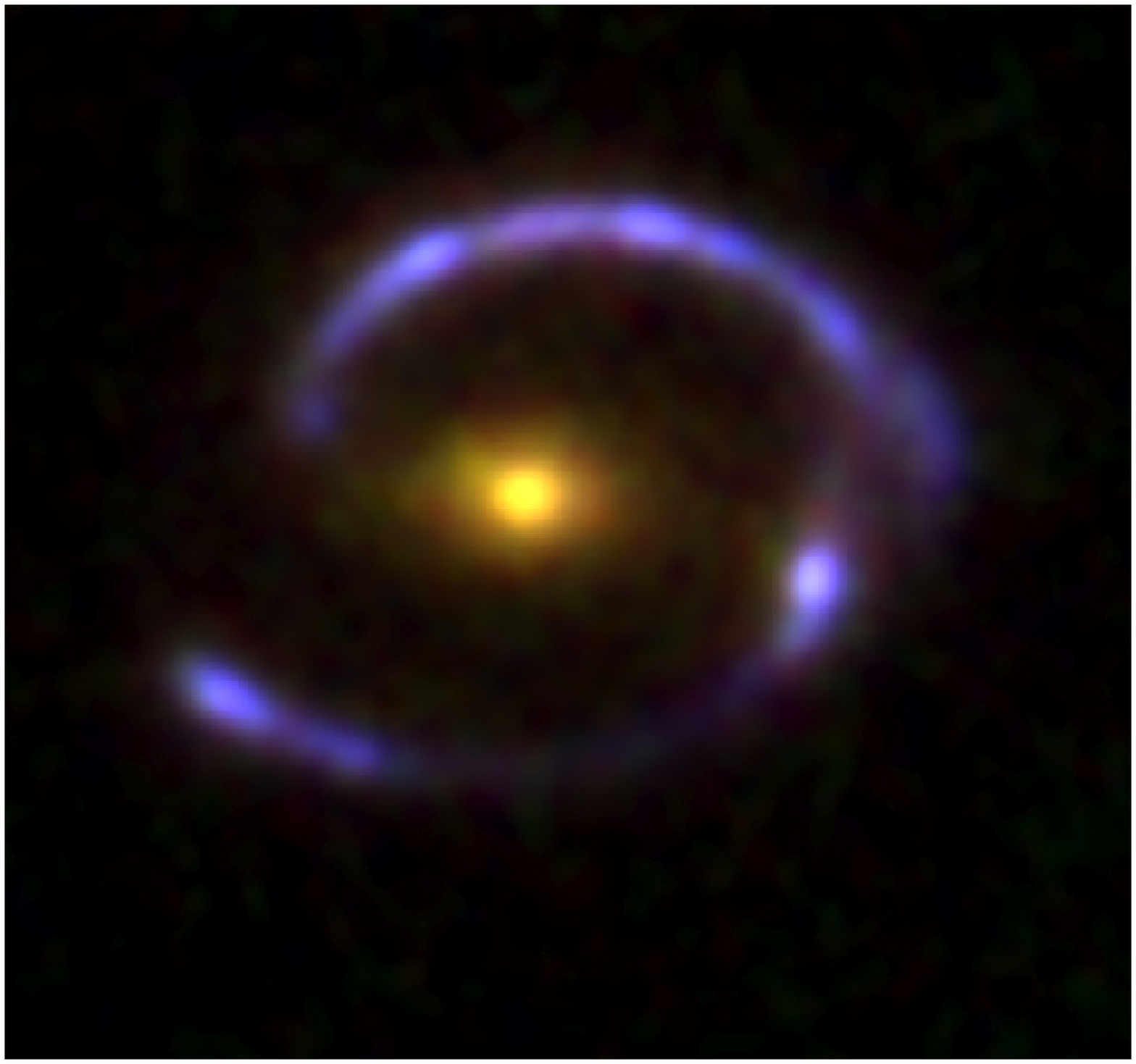}
\includegraphics[width = 210 pt, height=160 pt]{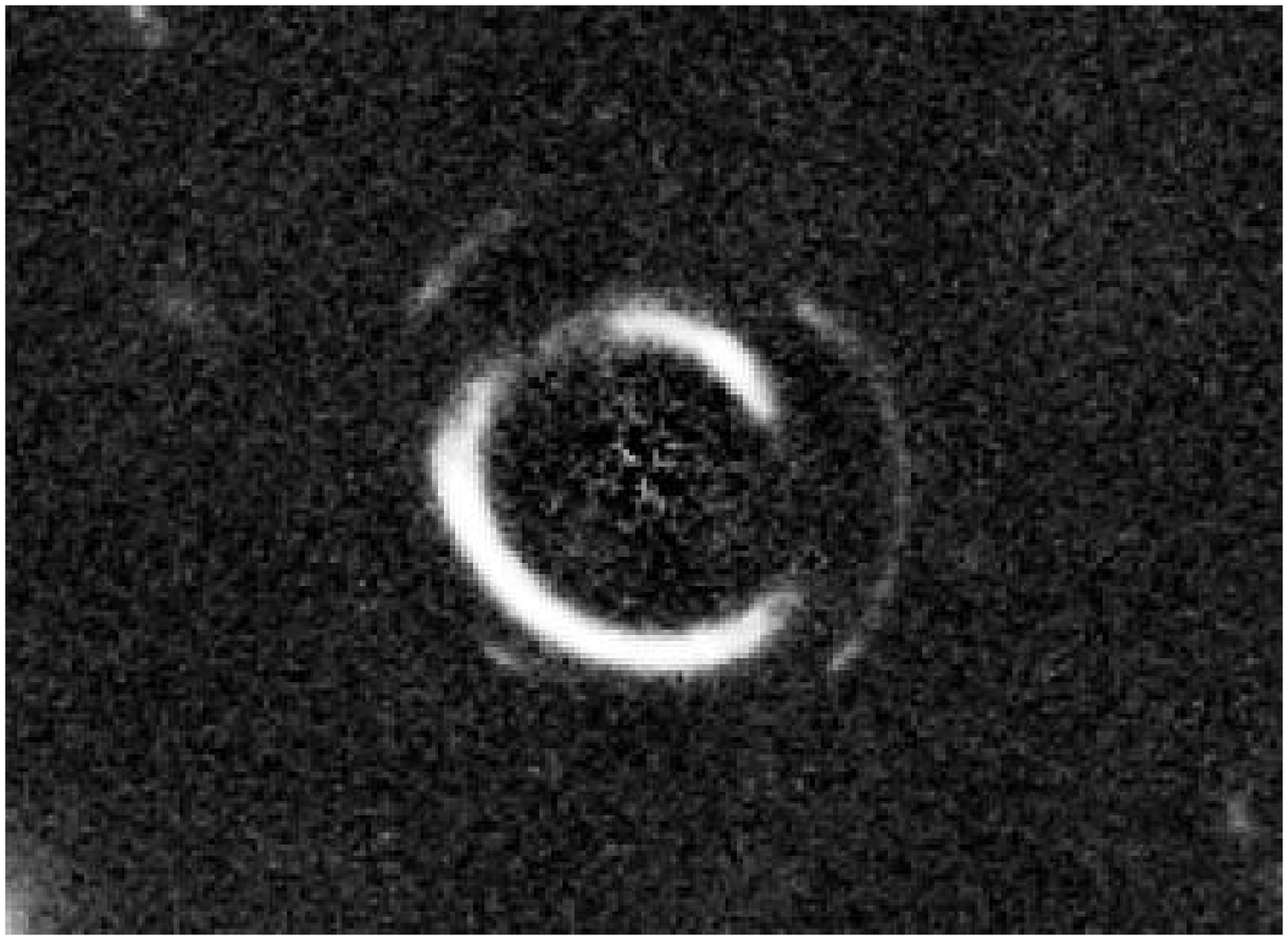}
\end{flushright}
\caption{\textit{Left:} The ``cosmic eye'' elliptical galaxy lens \cite{stark08}. If it is assumed that the baryonic
component of mass in the yellow lens galaxy traces the  distribution of light, the distribution of dark matter must be
cuspy, in line with predictions from simulations \cite{smail07,dye07}. \textit{Right:} a rare alignment of two sources behind one 
elliptical galaxy SDSSJ0946+1006 (after subtracting the foreground galaxy) \cite{gavazzi08}. This produced concentric, near-perfect
Einstein rings, and allows particularly precise measurement of the radial slope of the mass distribution, 
since both rings constrain the total interior mass. Both of these high
resolution images were obtained with the Hubble Space Telescope, but the second system was initially found in unresolved ground-based
data, from the spectroscopic signature of two sources at different redshifts \cite{bolton04}.}\label{fig:eye}
\end{figure}

Debate has been raging for several years about the inner profile of dark matter haloes within the scale radius. Examples
of elliptical galaxies
shown in figure~\ref{fig:eye} have been found with a cuspy $\beta\approx 1$ inner dark matter component, as
expected from the original NFW simulations. These include the ``jackpot'' double Einstein ring, which provides the
best lensing-only galaxy mass profile, and also shows that a dark halo is required.
However, most observations tend to prefer a flatter profile,
with a low $\beta\ls0.5$ on all mass scales. For example, the kinematics of dwarf and low surface brightness
galaxies, which are expected to be dark matter dominated throughout, suggest a dark matter distribution closer to a ``core''
($\beta=0$) and incompatible with NFW haloes \cite{simon03,kuzio08}. Rotation curves of spiral galaxies also indicate
lower central densities than expected in NFW haloes \cite{persic96,navarro00,binney01,gentile05}. The discrepancy is fundamentally
interesting because it could be due to the properties of dark matter: either the free-streaming of a partial `warm'
component away from the gravitational potential well, or a low level of (self-)interaction via the weak force to
provide pressure support and thus prevent collapse \cite{spergel00,moore02}. However, neither provides a complete
explanation. For example, it has been pointed out that the self-annihilation of dark matter particles would only
provide significant heating in dense cores with cuspy slopes $\beta\gtrsim1.5$, even assuming a generous interaction
cross-section and efficient transfer of released energy into the baryons \cite{natarajan08a}. Slopes with $\beta=1$
provide a factor of $1000$ less heating.

Galaxies that can be further studied because they act as strong gravitataional lenses appear to be a representative
sample of all galaxies -- chosen solely by the chance of another source lying directly behind them. Two independent
studies confirm that the fraction of elliptical galaxies acting as a lens is independent of local environment and stellar
properties \cite{faure09,treu09}. More elliptical galaxies overall are found in clusters, and simulations predicted that
these would have a lower stellar to total mass ratio, and more lensed images with larger arc radii \cite{hilbert07}.
However, the encouraging observational result confirms that, at least within current statistical limits, the lens galaxy
selection have the same gross characteristics (e.g.\ inner profile slope, which increases the strong lensing cross
section) as those in the field, and there is negligible boost from additional mass in large-scale structure.

The main challenge in extending the interpretation of the total mass near the centres of haloes will be that most of that mass is baryonic,
and therefore influenced by additional astrophysical processes that require more work to be understood and subtracted \cite{nipoti09}. 
Observationally, galaxies appear sharply divided into (mainly faint, discy and rapidly rotating)
morphologies with a central cusp, and (mainly brighter, boxy and slowly rotating) morphologies with a central core
\cite{lauer07}. Cored galaxies also tend to exhibit radio-loud active nuclei \cite{balmaverde06} and X-ray
emission \cite{pellegrini05}, confirming that major merger events, especially the merging of binary black holes,
profoundly impact the central distribution of baryons. 
Simulations demonstrate that baryonic cusps can be gradually softened by stellar winds or supernovae. 
Dynamical friction can build up a central baryonic cusp, but preserve the overall mass distribution by softening the 
dark matter core even if $0.01$\% the total mass exists in subclumps \cite{elzant01,elzant04}. 
The dark matter profile can also be softened by scattering from or accretion onto central black holes, especially when
tidally stirred by infalling satellites \cite{zhao02,read03}. Any remaining mass is rearranged into flat
$(\beta\approx0)$ cores particularly efficiently in low-mass dwarf galaxies or if the mass loss events are
intermittent \cite{read05}. 

Massive clusters of galaxies provide more multiple strong lensing systems, including radial arcs, and are  marginally
better understood.  Since cuspy cores act as more efficient gravitational lenses, the inner slopes of haloes can also be
statistically constrained by the abundance on the sky of any strongly lensed arcs \cite{wyithe01,ma03}, the relative
abundance of radial to tangential arcs \cite{molikawa01}, and the relative abundance of double- to quad-imaged systems
\cite{oguri04,mandelbaum09}. The high resolution of space-based imaging is particularly needed to identify radial arcs,
which are typically embedded within the light from a bright central galaxy. A systematic search of $128$ clusters in
the Hubble Space Telescope archive \cite{sand05} found a uniform ratio of $12$:$104$ radial to tangential arcs with
large radii behind clusters of a wide range of mass. This ratio is consistent with $\beta<1.6$, although the
interpretation again depends upon the assumed mass of the central galaxies. 

Studies of individual galaxy clusters have been particularly effective when combining strong lensing measurements with
optical tracers of the velocities of stars within the central galaxy. Analysis of the cores of nearly round, apparently
``relaxed'' clusters initially found $\beta=0.52\pm0.05$ \cite{sand02,sand04}. At the time, this was thought to be low,
and concerns were raised about various simplifying assumptions \cite{dalal03,bartelmann04,meneghetti07}. A more
sophisticated analysis \cite{sand08} of two clusters has since reached similar conclusions. However, degeneracies were
noticed that could only be broken by mass tracers at larger radii, and weak lensing observations are currently being
used \cite{newman09} to extend the range of measurement.


Another interesting observable is the light travel time to multiple images of a strongly-lensed source, 
which may differ by months or years (in a cluster, or weeks or months in an individual galaxy) because of
variations in both the geometric distance and the accumulated amount of gravitational time dilation. If the source
is itself variable (e.g. a quasar with varying output), the relative travel time can be measured between the
images. This is usually used to determine the value of Hubble's constant, by assuming a mass distribution for the
lens, often through the shape of strong lensing arcs, which then specifies the relative distance and time dilation
along the two paths \cite{refsdal64,coe09,suyu09}. However, if cosmological parameters are externally known, the same technique
can be used to improve constraints on the inner profile \cite{dobke06}.

\subsection{Substructure}

\begin{figure}[!t]
\begin{flushright}
\includegraphics[width = 160 pt]{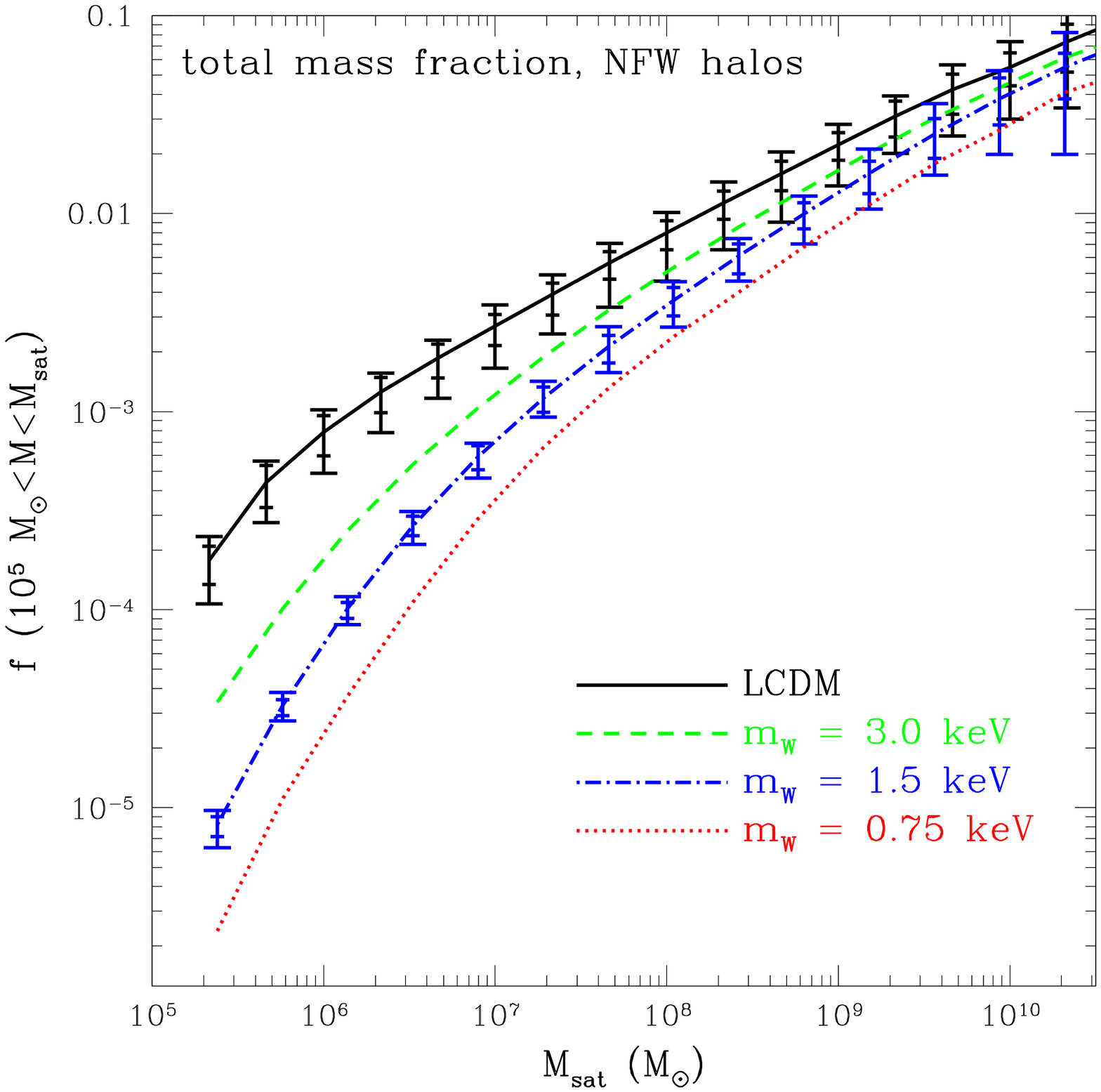}
\includegraphics[width = 230 pt]{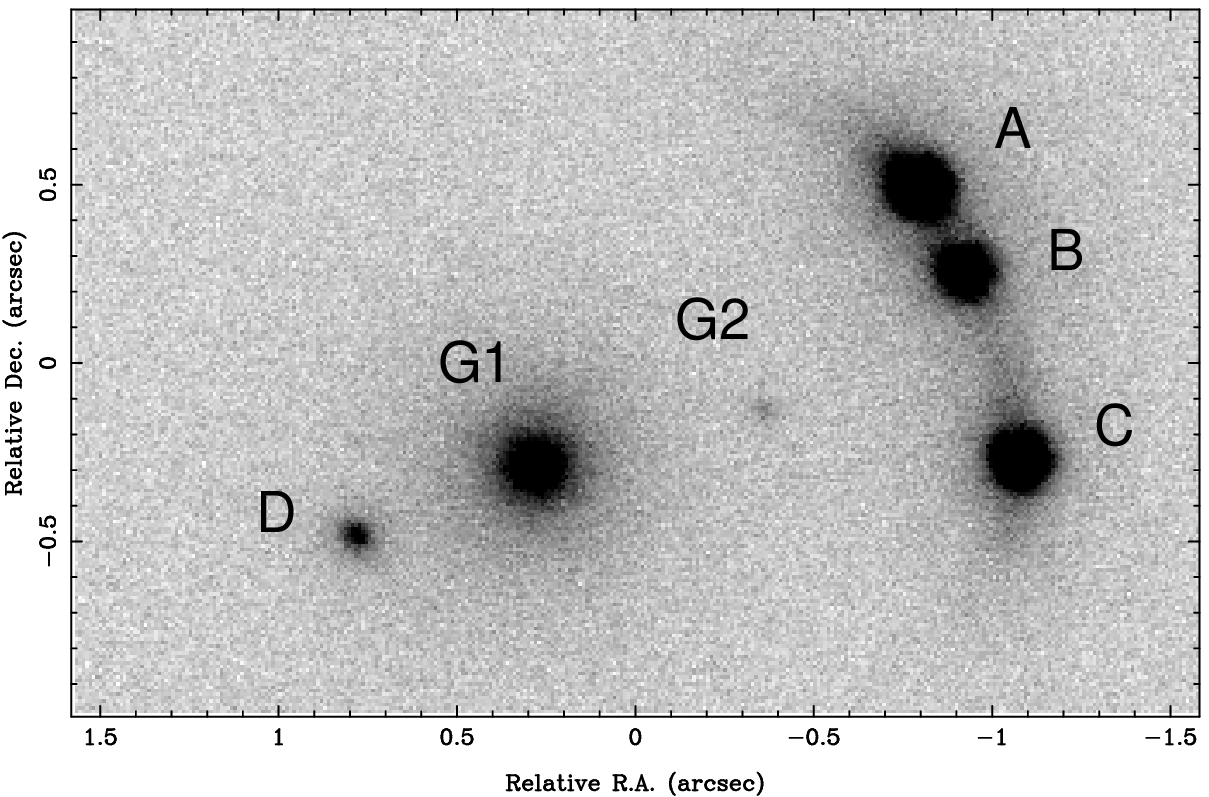}
\end{flushright}
\caption{Substructure in galaxies' dark matter haloes.
\textit{Left:} Expected fraction of mass in subhaloes more massive than $10^5~M_\odot$
from a semi-analytic model incorporating 
merger histories, survival probabilities and destruction rates of infalling substructure \cite{zentner03}.
Lines show models of standard cold dark matter (black) and warm dark matter with various particle masses 
(colours). Error bars reflect statistical scatter between realisations of the model calculations.
\textit{Right:}
Adaptive optics 2.2~$\mu$m imaging of lensed quasar B2045+265 \cite{mckean07}.
The lensed images (A, B, C, D) have one of the most extreme anomalous flux ratios known: models of lens galaxy
G1 require that B
should be the brightest image, but instead it is the faintest, suggesting the presence of an additional
perturbing mass. Indeed, this high resolution imaging revealed a small satellite galaxy (G2) that explains the
anomalous flux ratio.}\label{fig:qsolens}
\end{figure}

Structures in the Universe grow hierarchically via the ingestion of progressively larger objects, each of which themselves grew
from the merging of smaller units. Not all of the material in subhaloes is stripped from merging subhaloes and
smoothly redistributed amongst the cluster. Speculation still continues
\cite{calcaneo00,moore01,madau08,diemand08} on the expected fraction of mass in substructure, with estimates
typically ranging from $\sim5$--$65\%$ and options for the relatives sizes of the substructure shown in the left
panel of figure~\ref{fig:qsolens}. According to semi-analytic models, the substructure mass fraction varies with
cluster age and assembly history, so ``archaeological'' investigation can probe dark matter physics during
mergers, and also ``age-date'' smooth haloes that formed early and grew little, versus clumpy haloes that formed 
in a recent flurry of activity \cite{smith05,delucia04,taylor04,smith08}. 

The largest dark matter simulations currently resolve up to four generations of vestigial haloes within haloes
\cite{springel08}. A robust prediction of several hundred first generation subhaloes around galaxies like the Milky Way
led to the well-known  ``missing satellites problem'', because too few were observed. This issue is beginning to be
resolved, as larger and deeper surveys have finally unearthed a wealth of faint dwarf satellite galaxies
\cite{strigari08}, each of which contains very few stars (see \S\ref{sec:quantity_gals}) but a lot of dark matter
\cite{kazantzidis04}. To fine-tune the number of these ``dark haloes'', the focus of the debate has now switched from the
nature of dark matter to astrophysical issues such as the efficiency of star formation and reionisation. Beneath the four
hierarchical generations of substructure, simulations also contain an additional smooth component of dark matter. This
may imply that the fourth level is the end of the self-similarity, or may just be a limitation of the simulation
resolution. Endless, fractal self-similarly in the substructure mass function is appealing because it implies a patchy
distribution of dark matter within the Milky Way \cite{kamionkowski08}. If the Earth is currently within a void, this
could explain the lack of robust evidence for local dark matter from subterranean direct detection experiments. However,
testing this theory is difficult because star formation is so strongly damped in small dark matter haloes that they can
only be indirectly observed. Once again, gravitational lensing provides the only tool to do this in structures outside
the Local Group, studies of which will soon become statistically limited by the small sample size available.

Strong gravitational lensing is most sensitive to small mass variations close to a line of sight, and multiple images in which
light followed paths through the lens separated by more than the scale of substructure provide a unique
opportunity to observe a source both with and without its effect. The smooth potential of a galaxy
lens affects all the multiple images in a coherent way, but substructure affects each image individually. 
Thus the amount and mass function of substructure can be probed in the relative positions, fluxes and time delay
of multiple images of (point-like) quasar sources  \cite{chen07,williams08,keeton09}, and also ``fine structure''
variations in (extended) galaxy sources \cite{inoue05}.
Caution must be taken to rule out discrepancies caused by differential light propagation effects along alternative
paths through the lens, and (for optical imaging of quasar sources) potential confusion from month--year long
microlensing events by stars in the lens galaxy \cite{kochanek03}. In both cases, multiwavelength imaging provides
the quickest solution. Firstly, infra-red light is much less absorbed or scattered.
Multiwavelength imaging of lensed quasars is especially useful because the intrinsic source image ahanges
size at different wavelengths. A coarse model of the mass distribution in the lens
can be determined from multiple imaging of the large ($\sim1$~kiloparsec), low energy narrow line emission region.
Smaller substructure in the lens introduces millilensing (small Einstein radius lensing) that perturbs the relative 
intensity or ``flux ratios'' of the same multiple images, as seen in higher energy radiation originating from the 
$\sim 1$~parsec broad line region  \cite{moustakas02}. The fraction of mass in
stars within the lens can be found from ``microlensing densitometry'' variations of the continuum emission from 
the $\sim 10^{-3}$~pc central accretion disc \cite{morgan08,pooley08,minezaki09}.

Substructure of $10^{4}-10^{7}~M_\odot$ in elliptical galaxies provides the only explanation for observations of
anomalous flux ratios in multiply imaged quasars \cite{kochanek03,minezaki09,dobler06}. Indeed, very deep
near-infrared imaging in one of these cases, shown in the right panel of figure~\ref{fig:qsolens}, did eventually
find emission from one of the large pieces of substructure. Curiously, the  amount of substructure implied by
millilensing seems to be greater than that predicted by simulations, turning the missing satellites problem on its
head \cite{more09,xu09}. Time delays from substructure lensing have also been tentatively detected as changes in
the order of arrival time of multiple images \cite{keeton09b}.

Within galaxy clusters, group-scale substructure was revealed from early ground-based studies of strong lensing
arcs \cite{pello91,kneib93,kneib95}, and subsequent Hubble Space Telescope imaging refined the precision
\cite{kneib96}. An sample of $10$ clusters observed with Hubble ($5$ of them including strong-lenses) demonstrated
that $(70\pm20)\%$ of clusters contain major substructure, including multiple main density peaks \cite{smith05}.
In a further sample of $20$ strong lensing clusters it has recently been confirmed that the amount of substructure
required in parametric mass reconstructions is indeed anticorrelated with the dynamical age of the cluster, as
measured by the contribution of the brightest cluster galaxy and the magnitude difference between the two
brightest central galaxies, as seen in near-infrared light \cite{richard09b}.

The fundamental limitation with strong lensing techniques is the finite number of sight lines through the Universe that
end in a strong lens. Weak gravitational lensing analyses can overcome this in the most massive clusters. The weak
lensing signal is sensitive to all mass within an extended radius of the line of sight, so the reconstruction of
the mass distribution is inevitably non-local, but individual $10^{11}-10^{12.5}~M_\odot$ subhaloes can be found
amongst the larger net signal from the host halo with HST imaging \cite{natarajan04}. A weak lensing analysis of
five massive clusters \cite{natarajan07} shows a significant signal of substructure, excluding the possibility of
an entirely smooth mass distribution, and with a mass function $N(M)$ of substructure that is noisy but consistent
with simulations. The substructure contributes $10$--$20$\% of the clusters' total mass. On even larger (and
entirely statistical) scales, the accumulated presence of small-scale substructure can move power to small scales
$k\gs100h$~Mpc$^{-1}$ in the matter power spectrum, from large scales $1\ls k\ls100h$~Mpc$^{-1}$ that should be
accessible with the next generation of dedicated cosmic shear surveys \cite{hagan05}.

Flexion mapping is the most exciting compromise between these two regimes. The flexion signal is more local than
weak lensing and sensitive to the gradient of the mass distribution, which can be large near substructure even
when the total amount of mass is not \cite{bacon05,leonard09,shapiro09}. A direct reconstruction of the mass
distribution from flexion has been achieved in cluster Abell~$1689$ \cite{leonard07,okura08}, and revealed a new
subclump not resolved by weak lensing. The amount of substructure in dark matter haloes down to masses of
$\sim10^9~M_{\odot}$ should also be statistically detectable as an excess variance in the flexion signal
\cite{bacon09}. The flexion signal has a (small and spatially uniform) component of variance due to the intrinsic
shapes of distant galaxies. This is unaffected by a cluster with a smooth distribution of mass
\cite{lasky09,hawken}, but any substructure will increase this variance, which can be measured in radial apertures
about a cluster centre, and averaged over many clusters to overcome noise.

\section{Properties of dark matter} \label{sec:properties}

\subsection{Gravitational interaction} \label{sec:gravity}

It is assumed in the default cosmological scenario that dark matter interacts via normal Einstein gravity. The
ubiquitous coincidence of mass and light on all scales \cite{kneib03,gavazzi04,massey07a,limousin09} certainly
demonstrates that dark matter and baryons are mutually attracted. But the best probe of dark matter's precise
interaction lies in the tidal gravitational stripping of dark matter subhaloes as they are accreted on to a larger
structure. High resolution simulations of galaxy clusters \cite{ghigna00,gao04} show that tidal stripping should
rapidly reduce the mass of dark matter subhaloes as they are accreted on to a larger structure, compared to both
isolated haloes and also those already at the centre, which are only weakly affected by tidal forces. The
stripped dark matter is dispersed into the smooth underlying distribution.

\begin{figure}[t]
\begin{flushright}
\includegraphics[width=250 pt]{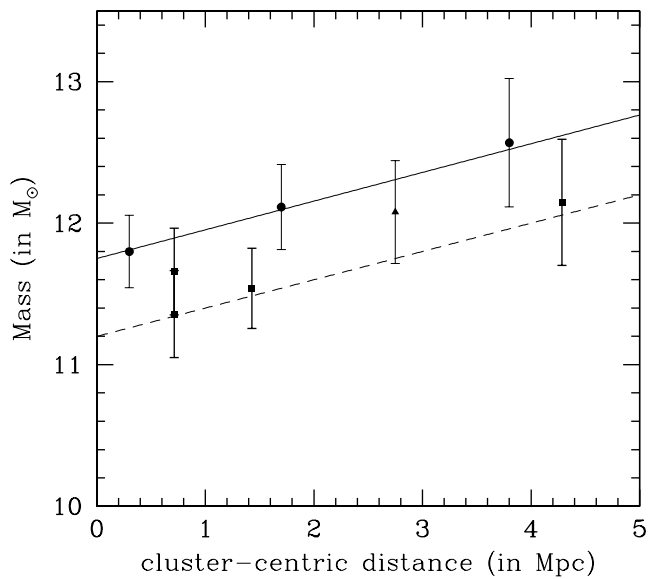}~~~~~~~~~~~~~~~~
\end{flushright}
\caption{
The removal of mass from the dark matter haloes around elliptical galaxies of a fixed luminosity by tidal
gravitational forces, as they fall into a large cluster \cite{natarajan09}. Circles and triangles mark
observations from gravitational lensing data, and squares show predictions from the Millenium $n$-body
simulation. The radial trend is consistent with dark matter being stripped as expected by a full
gravitational force; the different normalisations reflect uncertainty in the overall mass to light ratio.
}\label{fig:truncation}
\end{figure}

Observational evidence confirms this scenario, although current statistical uncertainties are large, 
since only the most massive clusters have substructure that can be studied in detail, and
homogeneous samples have only been gathered for a few of these. Nevertheless, 
weak lensing measurements from CFHT show that typical
elliptical galaxies of a given brightness live inside haloes that extend to $377\pm60$~kpc
\cite{hoekstra04}. Independent measurements of galaxies with a different camera on CFHT yield a
size of $\sim430$~kpc in the outskirts of supercluster MS~$0302$+$17$, but galaxies near the cluster
core are truncated at a radius $\ls290$~kpc \cite{gavazzi04}. Higher resolution weak lensing measurements are not
possible from the ground because of the smoothing required to achieve statistical precision, but an
even more thorough treatment has been made of galaxy cluster Cl~$0024$+$1654$, which has been observed
extensively from HST. The morphologies of the infalling galaxies appear to change dramatically only once they
fall within $\sim 1$~Mpc of the cluster core \cite{treu03}. However, as shown in figure~\ref{fig:truncation}, the weak lensing signal around elliptical galaxies
within $3$~Mpc indicates truncated haloes with a mean total mass of $1.3\pm0.8\times10^{12}~M_\odot$, compared
to $3.7\pm1.4\times10^{12}~M_\odot$ for similarly bright galaxies in the outskirts \cite{natarajan09}.
Indeed, different physical effects seem to dominate in three distinct zones around a cluster
\cite{treu03,natarajan09}. Outside the cluster virial radius, galaxies are entering the cluster for the first
time, and evolution is driven by the rare interactions and mergers of individual objects \cite{ghigna98}.
Inside a transition region of a few megaparsecs, tidal stripping of dark matter haloes begins to decrease the
mass-to-light ratio of galaxies that may make several passes through the cluster core. Within the central
megaparsec, tidal stripping is strongest, but baryonic effects including ram-pressure stripping also affect
the observed morphology of galaxies' baryonic component. There is slight evidence that tidal stripping in
Cl~$0024$+$1654$ is less efficient than in dark-matter only simulations \cite{natarajan09}, but this is probably
due to additional baryonic effects.

Measurements of strong lensing can push the analysis down to even smaller (kiloparsec) scales. Parametric
reconstructions of the haloes of elliptical galaxies
\cite{natarajan98,natarajan02,limousin07,halkola07,treu09} confirm that cluster members have less massive dark
matter haloes than galaxies in the field, and are truncated at radii around $17$-$66$~kpc, depending on the
cluster. There is also preliminary evidence that the inner profiles of galaxy dark matter haloes are
steepened by gravitational tidal effects during infall \cite{treu09}. This is expected from simulations \cite{dobke07}, and
also needed to resolve discrepancies between measurements of Hubble's constant from time delays
in strongly lensed multiple images behind a foreground cluster, and those derived from independent
techniques. With larger surveys planned in the future, this technique promises to be very fruitful in
constraining the properties of dark matter. 

%


It is particularly important to pin down the interaction of dark matter with gravity because all the current
evidence for dark matter is gravitational. Indeed, many attempts have been made to circumvent the need for dark
matter by modifying general relativity. Some of these theories of gravity can also predict the accelerating
expansion of the Universe, otherwise attributed to dark energy \cite{song07,hu07a,hu07b,camera09}. Modifications
of general relativity generally involve additional source terms (e.g.\ a scalar field) to explain individual
phenomena, such as the fast rotation of galaxies or the separation between X-ray and lensing signals in the bullet
cluster \cite{bekenstein04,zhao06}. However, none of these alternative theories has yet been able to consistently
explain the whole range of dark matter observations that are successfully verified within the standard
cosmological model, without requiring at least a small additional component of weakly interacting mass
\cite{hoekstra04,natarajan08,knebe09,tian09}. 

\subsection{Electroweak interaction} \label{sec:crosssection}

\begin{figure}[t]
\begin{flushright}
\includegraphics[scale = 0.33]{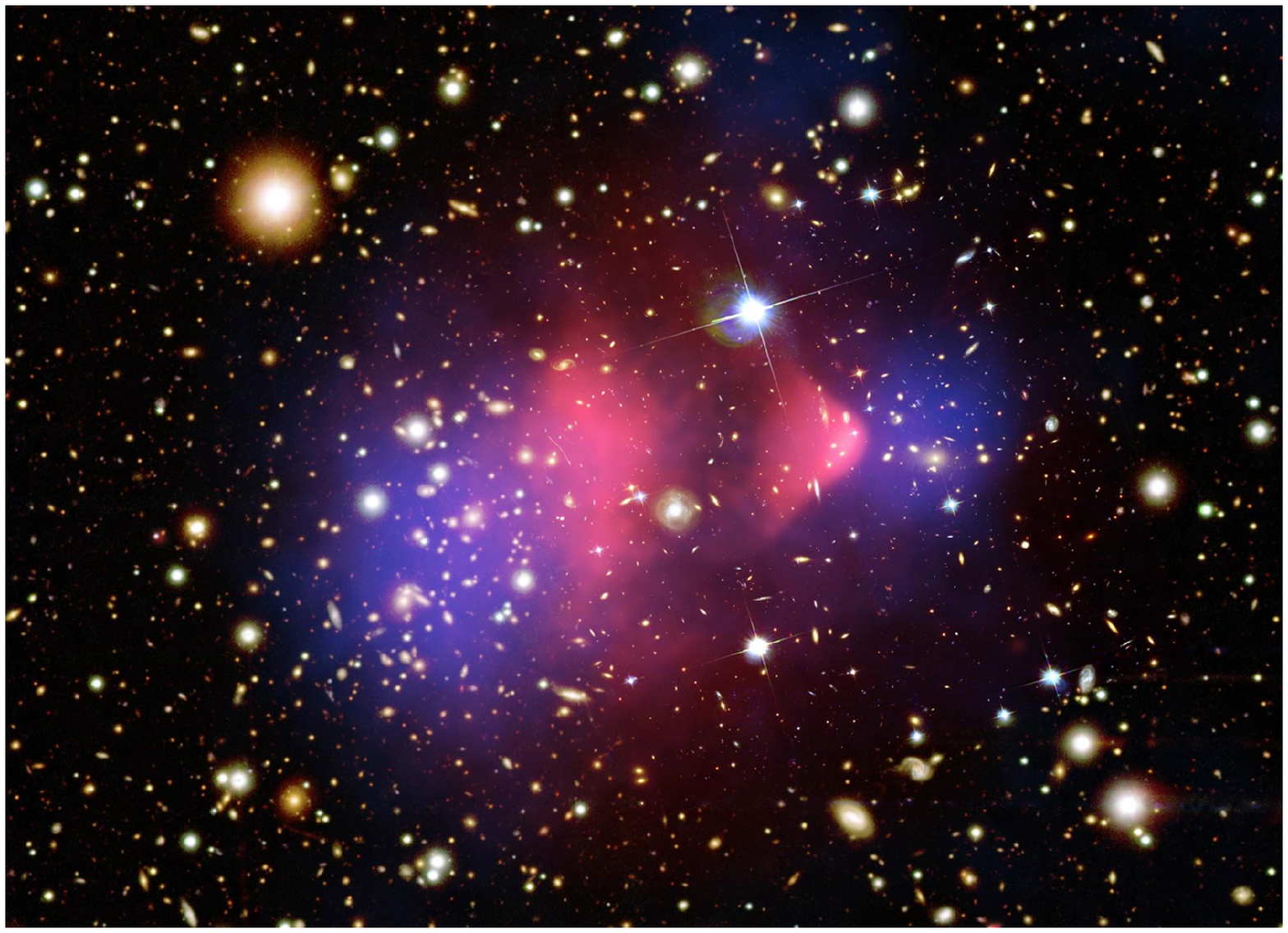}
\includegraphics[scale = 0.33]{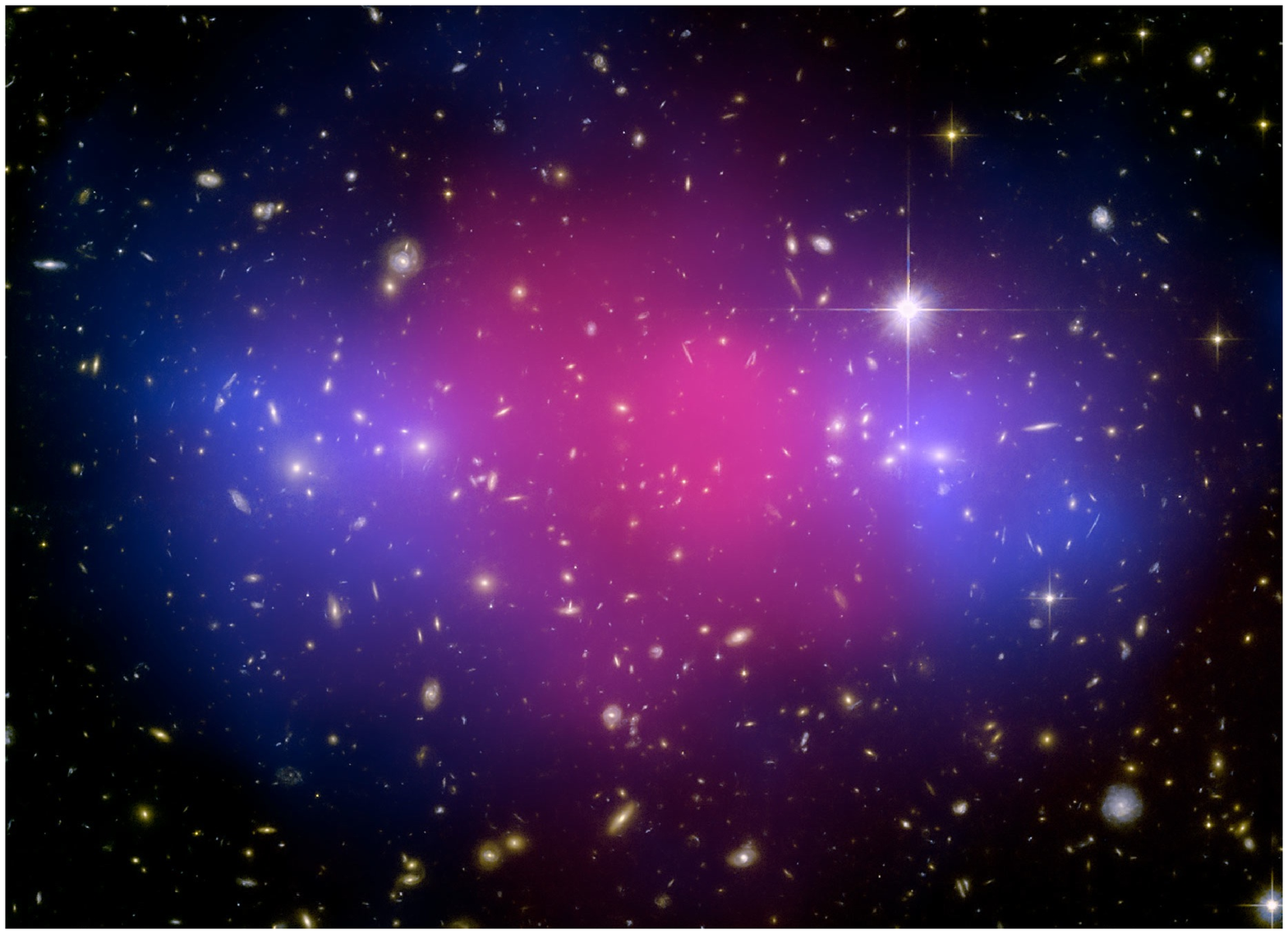}
\end{flushright}
\caption{
The ``bullet cluster'' 1E0657-56 and ``baby bullet'' MACSJ0025.4-1222. The
background images show the location of galaxies, with most of the larger yellow 
galaxies associated with one of the clusters. The overlaid pink features show
X-ray emission from hot, intra-cluster gas. Galaxies and gas are baryonic
material. The overlaid blue features show a reconstruction of the total mass from
measurements of  gravitational lensing. This appears coincident with the
locations of the galaxies, implying it has a similarly small interaction
cross-section. However, there is far more mass than that present in the stars
within those galaxies, providing strong  evidence for the existence of an
additional reserve of dark matter. (Figure credit: Left: X-ray:
NASA/CXC/CfA/ M.Markevitch et al.; Lensing Map: NASA/STScI; ESO WFI;
Magellan/U.Arizona/ D.Clowe et al.\  Optical image: NASA/STScI; 
Magellan/U.Arizona/D.Clowe et al.; Right: NASA/ESA/M.Bradac et al.).}
\label{fig:bullet} \end{figure}

The complex physical processes during the assembly of clusters from subhaloes
stir up the distribution of baryonic mass and obscure much of the
behaviour of dark matter. However, the differences between
baryons and dark matter are also highlighted by their different reactions to these processes.  The most
striking example of this, and the cluster that has provided the 
most direct empirical evidence for dark matter, is undoubtedly the ``bullet cluster''
1E~0657-56 \cite{markevitch02,clowe04,clowe06}, shown in figure~\ref{fig:bullet}.

Galaxy clusters contain three basic ingredients: galaxies, intra-cluster gas, and dark matter. 
Inconveniently for 
those trying to interpret the total mass distribution, these ordinarily come to 
rest in approximately the same place.
However, this is not always the case. The bullet cluster is strictly two clusters that collided about $150$
million years ago,
within or close to the plane of the sky, and the ingredients have become separated. Rather like the scattered
ejections from a particle collider, the trajectories of the cluster detritus are governed by the properties
of its components. Since individual galaxies within the clusters (and stars within those galaxies) are well-spaced, they have a very low
collisional cross-section: most continued moving during the collision, and today lie far from the point of
impact. On the other hand, intra-cluster gas was uniformly spread throughout the incident clusters. This
had a large interaction cross-section and was slowed dramatically during the
collision. The two concentrations of hot gas, seen in X-ray emission, have now passed through each other, but
have not travelled far from the point of impact. The collision speed and gas density were sufficient for a
shock front to be observed in the gas from the smaller of the two clusters, allowing the determination of the
collision speed. Initial concern that the shock velocity of $4700$~km/s is higher than that expected for any
merging haloes \cite{koda08} has now been resolved by properly taking into account the relative motion of the
gas clouds in the rest frame of the smaller cluster.

Crucially, gravitational lensing observations require the third ingredient of the bullet cluster. A great
deal of extra mass ($\sim30-40\times$ than that seen in the galaxies' stars) is located near the galaxies, and
$8\sigma$ away from the gas peaks. To have travelled so far, this mass must have a self-interaction collisional
cross section $\sigma/m<1.25$~cm$^2$g$^{-1}$ at $68$\% confidence (or $\sigma/m<0.7$~cm$^2$g$^{-1}$ under the
assumption that the mass to light ratio of the initial clusters was the same) \cite{randall08}. Note that
this does not include a constraint on the dark matter-baryon interaction cross section. 

The visible separation between the three ingredients of each cluster is temporary. Within another billion
years, the mutual gravitational attraction of the galaxies, gas and dark matter will have pulled them back
together, spiralling back into the usual configuration of baryonic material within a dark matter cocoon.
Consequently, such objects are rare. One very similar ``baby bullet'' object has been detected
\cite{bradac08} (see right panel of figure~\ref{fig:bullet}), as well as a ``cosmic train wreck'' counter-example \cite{mahdavi07}, which shows separated
dark matter and gas components, but in a complex distribution that probably implies a collision between three
clusters. Collisions along the line of sight would provide complementary information, but the one possible
example \cite{jee07} is probably an artefact of spurious instrumental effects \cite{boffi07} 
and substructure within the cluster \cite{kneib03}. However, the race is now on to find more bullet clusters,
at a range of collision speeds, masses, impact parameters and angles to the line of sight \cite{shan09}. 
A statistical analysis of many bullet clusters would overcome uncertainties in any individual system, and
help dispel any lingering doubts that a set of chance effects (or features of the nature of gravity) conspire to
produce the observed appearance.

\subsection{Self-annihilation/decay}

As discussed in \S\ref{sec:cuspcore}, gravitational lensing observations of flat cores in the centres of dark matter
haloes could be explained by a finite self-interaction dark matter cross section. Relying on this effect to produce
cores in galaxy haloes requires a
cross section $\sigma\sim 0.56$-$5.6$~cm$^2$g$^{-1}$ \cite{dave01,ahn03}. However, smoothing the mass profile of
clusters requires a much larger $\sigma\sim 200$~cm$^2$g$^{-1}$ \cite{ahn05}. If both are to be explained by
self-interaction, either the
cross-section is velocity dependent and/or other astrophysical effects are dominant. 
Self-interacting dark
matter also generally produces haloes that are more spherical than standard CDM, especially in the core
\cite{dave01}, although current measurement uncertainty is too large to discriminate. Measurements from the
PAMELA satellite \cite{pamela}, ATIC balloon \cite{atic} and of the WMAP haze \cite{hooper07b,cumberbatch09}
tentatively suggest a high value of the related dark matter-dark matter annihilation rate. These results are
being greeted with cautious skepticism, as similar effects have not been reproduced in other detectors that
ought to see decay products.

If axions 
exist as a component of dark matter,
they would couple to standard-model particles \cite{baluni79,crewther79}
and should be detectable via photon-photon decay into a 
single optical emission line with a flux proportional to the dark matter density. 
Integral field spectroscopic searches for this signature emission from the dark matter haloes of two galaxy clusters
have benefitted from strong gravitational lensing \cite{grin07}.
The search efficiency was improved by a factor of $3$ over previous analyses
by correlating the search with strong lensing mass maps of the densest regions, 
where the emission from such decays is expected to be strongest. 
The sensitivity for emission line detection in the optical 
wavelength range allowed them to derive interesting upper limits on the two-photon 
coupling in the mass range $4.5$ eV$<m<7.7$ eV. This work highlights the potential of 
spectroscopy coupled with accurate maps of the dark matter distribution to explore a 
larger axion mass window at higher sensitivity, and the same data can also be used to constrain the decay rate
of other $\sim$5~eV relics, such as sterile neutrinos.

\subsection{Particles or planets?} \label{sec:macho}

In the broadest sense, ``dark matter'' refers to any matter that is undetectable through either emission or
absorption in the electromagnetic spectrum. Some astrophysical objects are naturally dark, such as black holes ---
and we face technological limitations in the detection of any faint sources. It is a particularly attractive
proposition that a large population of ``dark'', $\sim$planet-sized Massive Compact Halo Objects (MACHOs)
contribute a major component of the missing mass but have so far evaded detection. Such objects could be baryonic
in nature and could even fit within our standard understanding of stellar evolution, thus removing the need for any
new particles.

There have been extensive and sustained efforts to characterise the number of MACHOs in the halo of the Milky Way,
its satellites the Large and Small Magellanic Clouds,
and our neighbouring galaxy Andromeda (M31). Even though MACHOs are not visible themselves, whenever one passes in
front of a star its gravitational microlensing briefly brightens the
star. Since the volume of space along lines of sight that would cause microlensing is tiny, many millions of stars
need to be continually monitored.

Looking towards $12$ million stars in the Magellanic Clouds for $5.7$ years, the MACHO survey \cite{alcock00} found only
$13$--$17$ microlensing  events (and some of these have been challenged as supernovae or variable stars). At $95\%$
confidence, this rules out a model in which all of the Milky Way's dark matter halo is (uniformly distributed) MACHOs.
However, if all events are real, the rate is still $\sim 3$ times larger than that expected from a purely stellar
population, indicating either that they contribute up to $20\%$ of the Milky Way halo's mass \cite{popowski03}, or
a larger fraction of the Magellanic Cloud halo, in less massive bodies \cite{novati06}.
Also looking towards the Magellanic Clouds, the Exp\'erience pour la Recherche d'Objets Sombres (EROS) project 
\cite{tisserand07} found only $1$
event in $6.7$~years of monitoring $7$ million stars, compared to the $39$ expected were local dark matter 
composed entirely of $0.6\times 10^{-7}$--$15~M_{\odot}$ MACHOs. Looking towards the Magellanic Clouds and the densely populated
central bulge of the Milky Way, the Optical Gravitational Lensing Experiment
(OGLE) \cite{novati09,wyz09,udalski09} detected only $2$ microlensing events in $16$~years, and even these events are
consistent with self-lensing by stars, rather than MACHOs \cite{zhao99,novati09}. 
The OGLE results conclude that at most $19\%$ of the mass of the Milky Way halo is in
objects of more than $0.4$ $M_{\odot}$, and that at most $10\%$ is in objects of $0.01$--$0.2~M_{\odot}$. The
POINT-AGAPE experiment \cite{novati05,novati09b} observed unresolved (pixel) microlensing in the more distant Andromeda galaxy, 
and found that at most $20\%$
of its dark matter halo is in  $0.5$--$1.0~M_{\odot}$ mass objects (at $95$\% confidence).

The gravitational microlensing results provide convincing evidence that less than $20\%$ of the mass of the
Galactic halo is in the form of MACHOs. Coupled with cosmological measurements of the baryonic and total mass
density, this is strong evidence that remaining dark matter must be non-baryonic. Weakly interacting massive
particles have become well-established as the favourite candidate. However we note that while MACHOs down to planet
mass scales have been convincingly ruled out, the jump in scale from this to a sub-atomic description is not one
that should necessarily be taken for granted. 

The rarity of MACHOs has meant that every individual event has become a 
great source of information. An industry of follow up observations has developed to 
help try to determine the nature of 
massive compact objects as either neutron stars, black holes/dwarfs or something more
exotic.

\subsection{Warm and Hot Dark Matter} \label{sec:neutrino}

If dark matter is in thermal equilibrium, the mean velocity of the particles can be characterised by a dark matter
temperature. If the dark matter particles were produced in the early Universe by standard ``freeze-out''
from a primordial soup, this temperature is also related to the mass of the particles. 
Relativistic free-streaming of dark matter particles lighter than $\sim 1$~MeV would have
suppressed subsequent gravitational collapse. The large-scale structure visible today, and that in the
Cosmic Microwave Background \cite{Spergel}, require the dominant component of dark matter to have been dynamically cold
and therefore massive. However, it seems most likely that dark matter consists of several components,
potentially including a small contribution of dynamically warm or hot particle species such as gravitinos or
massive neutrinos. Massive Majorana ($\nu_{\alpha}=\bar\nu_{\alpha}$)\footnote{For cosmological purposes the
distinction between Majorana and Dirac neutrinos is not important. However, astrophysics could feasibly detect the
neutrinoless double beta decay possible with Majorana neutrinos, thus constriaining neutrino mass
(and distinguishing  between normal or inverted hierarchies).} neutrinos also provide an attractive theoretical
explanation for the baryon asymmetry \cite{Fukugita}. Note that, although very light, sterile neutrinos  
(which do not couple with any standard electroweak interaction) or axions
would be dynamically cold, since they were produced due to symmetry breaking and have never picked up large kinetic energy by 
being in thermal equilibrium with the Universe.


Neutrino particle physics and astronomy first joined forces to explain the 
``solar neutrino problem'', that fewer electron 
neutrinos than expected are detected from the Sun. The Solar Neutrino Observatory (SNO) 
\cite{Ahmed} explained this deficit as the oscillation of electron 
neutrinos to other ($\mu$ or $\tau$) flavours en route to Earth. SuperKamiokande \cite{Fukuda}
studied cosmic ray collisions and found a similar result that only $\sim 1/3$ the flux of muon neutrinos 
$\nu_{\mu}$ from cosmic ray collisions in 
the atmosphere were observed along the line-of-sight through the Earth. This implies an 
oscillation of $\nu_{\mu}$ to some other flavour with a scale length comparable to the radius of the Earth. 
Further evidence for neutrino oscillations 
comes from nuclear reactors (KamLAND, \cite{Eguchi}) and neutrino beam experiments 
(K2K, \cite{Ahn}). The LSND \cite{LSND} results suggest the existence of sterile neutrinos, and the 
MiniBOONE experiment \cite{MB} should confirm or refute this result. There have 
also been cosmological constraints on the abundance of sterile neutrinos \cite{Seljaks}. 

The rate of neutrino oscillations depends upon the (square of) the mass differences between neutrino flavours. 
Current constraints on the mass differences from large particle physics experiments are 
$|\delta m_{23}| \sim 0.05$~eV and 
$|\delta m_{12}| \sim 0.007$~eV. There are currently no strong constraints on 
$|\delta m_{13}|$, though upcoming experiments like T2K \cite{Itow} 
should measure this with an accuracy of $\sim0.05$~eV. 
Thus current constraints allow for two possible orders of the massive neutrino hierarchy: 
$m_1 < m_2 < m_3$ or the inverted hierarchy $m_3 < m_1 < m_2$. 
Particle physics experiments are planned that will measure the absolute mass scale via 
the beta decay of Tritium \cite{Mbeta,Mbetb,Mbetc,Mbetd,Main,Troi}. For example,
 KATRIN \cite{Osipowski} is expected to reach an accuracy of $\Delta m_{\nu_e}\sim 0.35$~eV. 
However, providing useful information on the hierarchy configuration will require 
an accuracy of $\Delta m_{\nu}<0.1$~eV \cite{Lesgourgues}.

If dark matter does include a component of massive neutrinos, they will have left a key signature in the Cosmic
Microwave Background \cite{Ma95} and in large-scale structure \cite{Eisenstein,viel05}. Neutrinos free-stream out of
gravitational potentials, reducing the amount of matter that can accumulate on small scales. This erases substructure in
dark matter haloes and softens their central cusps. This is observable via strong lensing of galaxies
\cite{spergel00,moore02} and quasars \cite{wdm1,wdm2}. The altered flexion variance signal on due to substructure could
detect $2$~keV neutrinos \cite{bacon09}.  

The suppression of small scale structure also shifts (free-streams) power onto the larger scales accessible by weak
lensing \cite{Hannestad06, Abazajian, Cooray}.  Current weak lensing observations of the cluster mass function, in
conjunction with the SDSS and 2dF galaxy redshift surveys, constrain the total neutrino mass  $m_{\nu}< 1.43$~eV (at
95\% confidence) \cite{knu}. Two independent analyses of cosmic shear in the CFHTLS Wide survey \cite{tereno,ichi} both
tentatively find an upper limit of  $m_{\nu}<0.54$~eV (at 95\% confidence) when combined with data from the CMB (WMAP 
3~year), supernovae and baryon acoustic oscillations. Whilst these constraints are heavily dependent on priors, the
statistical accuracy is well matched to that expected from the size of the survey. The next generation of large, three
dimensional cosmic shear surveys expect to detect warm dark matter particles with a mass of $\sim5$~keV \cite{mark09}
and constrain the effective number of neutrino species $N_\nu$ to $\sim0.1$ and $\Delta m_\nu$ to within a few times
$10^{-2}$~eV \cite{kitching08}. Weak lensing should even be able to distinguish between inverted and normal neutrino
hierarchies, because the lighter species become relativistic at a higher redshift than more massive particles and
consequently have a different impact on the dark matter power spectrum $P(k)$ \cite{debarn} and the weak lensing of the
CMB \cite{slosar}. 


%

\section{Future directions} \label{sec:future}

\subsection{Hardware}

\begin{figure}[t]
\begin{flushright}
\includegraphics[height=147 pt]{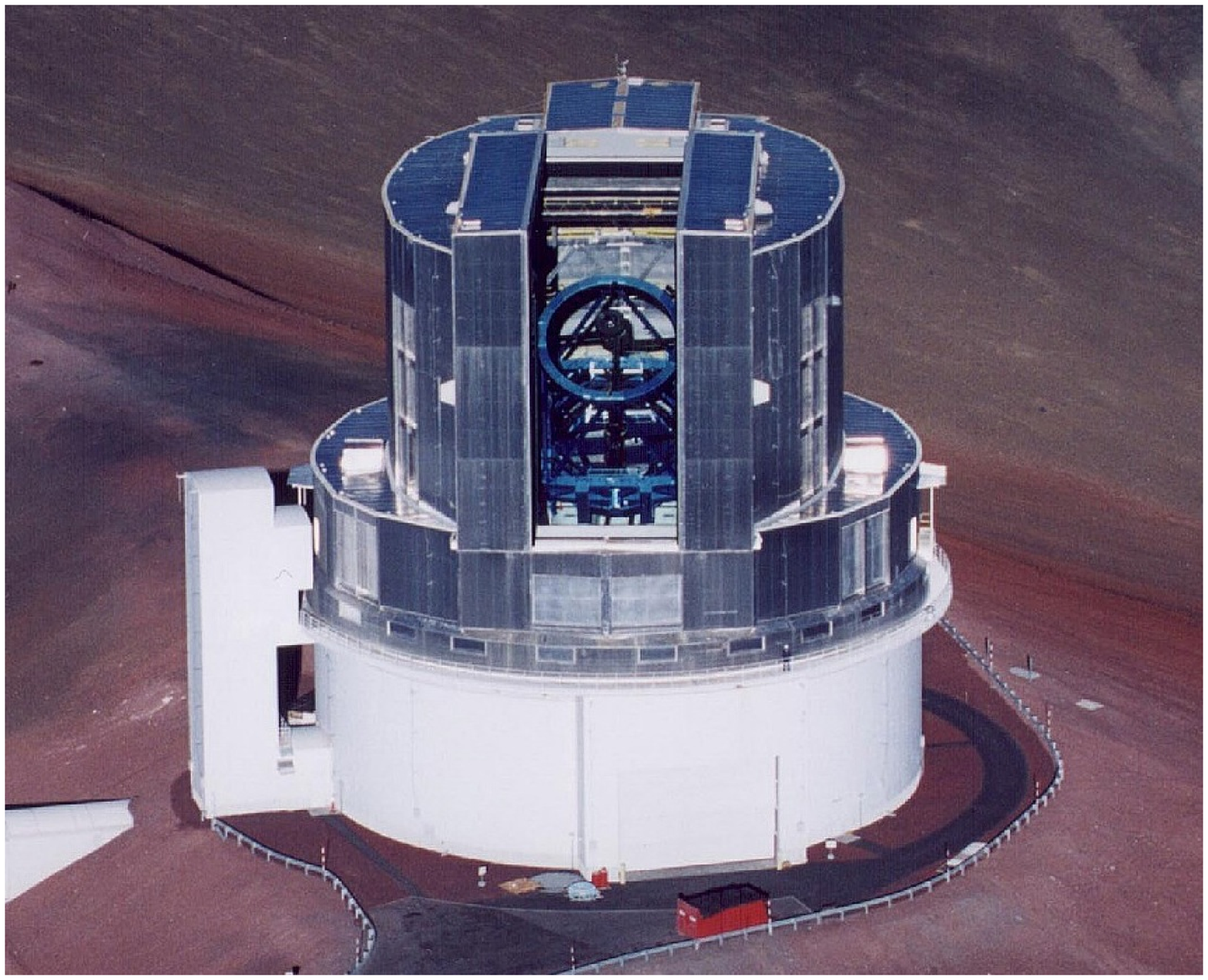}
\includegraphics[height=147 pt]{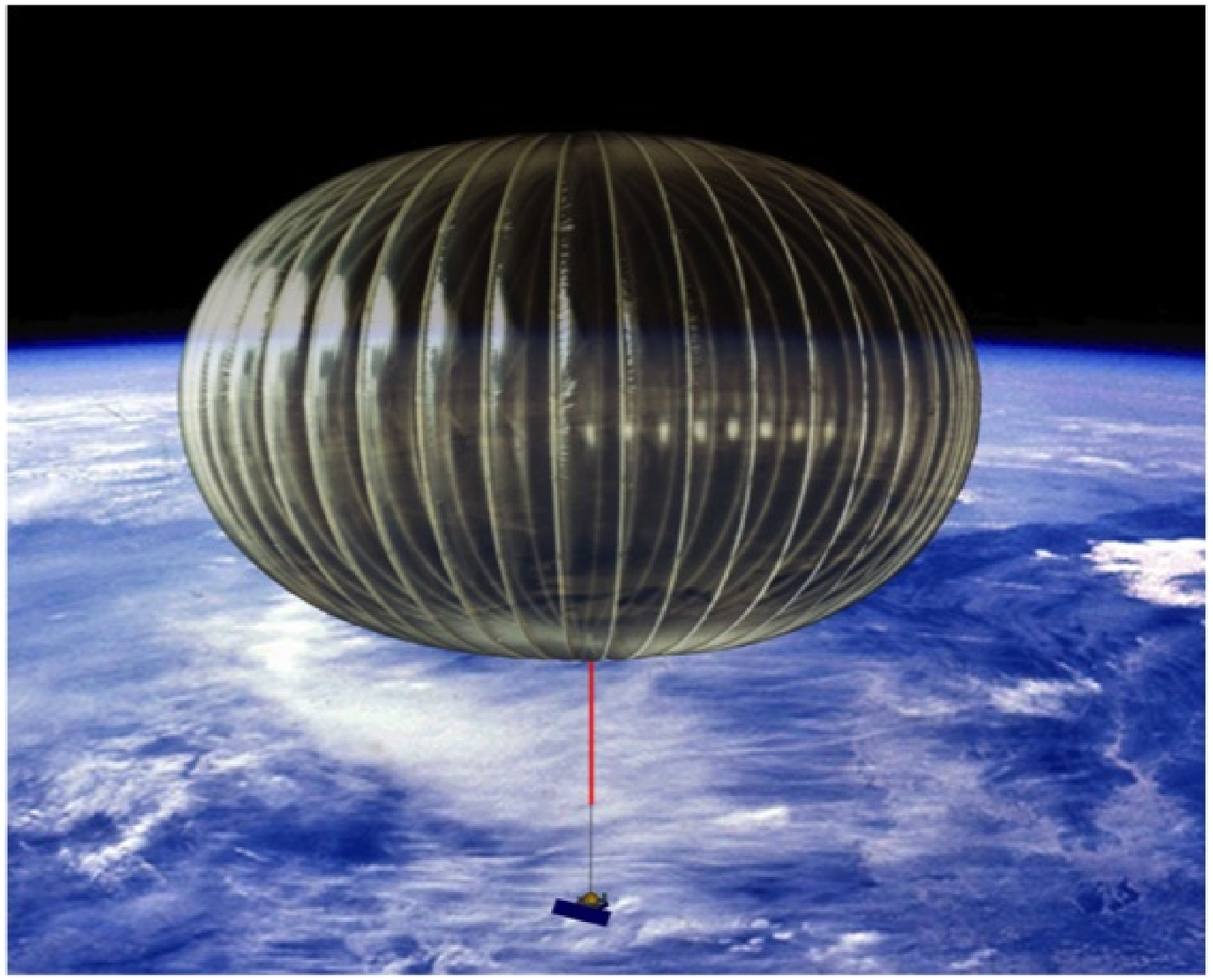}
\end{flushright}
\caption{The investigation of dark matter via gravitational lensing has a bright near-term future.
\textit{Left:} The 8m Subaru telescope atop the 4200m inactive volcano Mauna Kea, Hawaii.
Subaru will soon receive an upgraded imaging camera, mounted at prime focus -- the top of the telescope and 
visible in the photo. Prime focus allows a wide field of view, but requires sturdy support to avoid flexure of 
the optical path as the telescope is pointed in different directions around the sky (credit: National Astronomical Observatory of
Japan). \textit{Right:} Artist's impression of the planned HALO telescope, observing gravitational lensing 
from a long-duration balloon at an altitude of 35km, above 99\% of the Earth's atmosphere 
(credit: NASA Columbia Scientific Balloon Facility).}
\label{fig:future}
\end{figure}

Several facilities are being designed or converted for dedicated measurements of gravitational lensing.
The Japanese $8$m Subaru telescope on Mauna Kea, Hawaii, shown in the left panel of figure~\ref{fig:future}, was
built with weak gravitational lensing measurements explicitly in mind. 
Its robust construction stabilizes the optical path and has produced the highest quality imaging of
all ground-based telescopes. A replacement is currently being built for its $0.25$~square degree field of 
view SuPrime-Cam imager. The next generation, $1.5$~gigapixel
HyperSuPrime camera will have a $2.5$~square degree field of view \cite{miyazaki},
and will carry out a multiwavelength gravitational lensing survey covering $2000$~square degrees. The KIlo
Degree Survey (KIDS) should cover $1400$ square degrees,
exploiting a new, $0.3$~gigapixel, $1$~square degree OmegaCAM optical imager on the European $2.4$m VST telescope
at Paranal, Chile \cite{kids}. The Dark Energy Survey (DES) will cover $5000$~square degrees, using a $0.5$~gigapixel,
$2.2$~square degree DECam imager being designed for the $4$m Blanco telescope at Cerro Tololo, Chile \cite{lin}.
Most of this survey will also be covered by the South Pole Telescope's CMB survey to find galaxy clusters via the
Sunyaev-Zeldovich effect, allowing an important cross-correlation. Looking further ahead the Large
Synoptic Survey Telescope (LSST) is being planned, with a $6$m effective primary mirror and a
$3.2$~gigapixel optical camera \cite{lsst}. All of these surveys will include imaging at multiple optical and near-infrared
bands in order to obtain photometric estimates of galaxy redshifts and thus perform 3D lensing analyses.

Circumventing the drive toward ever-larger single-dish telescopes is the Panoramic Survey Telescope and
Rapid Response System (Pan-STARRS) on Haleakala, Hawaii \cite{kaiser02}. Survey imaging has just begun using a
single $1.8$m telescope (PS1) with a $1$~gigapixel optical camera to cover a shallow but full-hemisphere Wide
Survey plus a $70$~square degree Medium Deep Survey (to the same depth as the Canada-France-Hawaii telescope's deep legacy
survey, which covers only $1$ square degree). The detectors include the unique ability to transfer accumulated
charge between pixels during an exposure, following short-term variations in the atmosphere to improve image
quality. Most importantly, PS1 is a prototype for a much larger sequence
of facilities. Construction of PS2 has already begun: a duplicate unit that will be situated alongside and may be 
used in conjunction with the original. Its goal will be to demonstrate that repeat production of relatively small
telescopes can be cost-efficient. If this is successful, two more identical telescopes will be built to create PS4,
all potentially housed in the same dome but moved to a higher altitude site on Mauna Kea, and more still for PS16
situated at various locations around the world, allowing 24 hour-a-day continual observation of the sky \cite{kaiser04}. 

There have been some attempts to use radio data to measure the weak lensing effect \cite{patel09,chang04}. 
Such exploratory investigations suggest that surveys such as LOFAR and the Square Kilometer Array (SKA) could 
use long baseline interferometry (approximately $300$ to $500$~km for the SKA frequencies) 
to image radio sources at a sufficient resolution to measure weak lensing distortion \cite{ska,lofar}.
The expected sources are star-forming galaxies, whose spatial density is uncertain, but whose intrinsic shapes are probably
more regular than usual radio sources. If possible, such radio observations 
would also open up a whole new lensing regime: the large scale structure at very high 
redshift could be inferred from lensing of the pre-galactic Hydrogen $21$cm emission \cite{pilipenko07}.

The most exquisite weak lensing measurements require telescopes above the Earth's atmosphere. The Hubble Space
Telescope already offers imaging with more than ten times higher resolution --- and crucially stability -- than
even the Subaru telescope, although over a narrow field of view \cite{kasliwal08}. The proposed High Altitude
Lensing Observatory (HALO), illustrated in the right panel of figure~\ref{fig:future}, 
will fly on a long-duration balloon above $99\%$ of the atmosphere. If its pointing
accuracy can be successfully stabilized, it will offer wide-field imaging of almost space-based resolution for
about the same cost as most cameras currently under construction for ground-based telescopes. However, achieving the
ultimate precision in lensing observations of dark matter will eventually require deep imaging from a dedicated
space-bourne telescope in a dynamically and thermally stable environment at the Earth-Moon Lagrange point L2. The
SuperNova/Acceleration Probe (SNAP) and Dark Energy Space Telescope (DESTINY) mission concepts proposed for the
NASA/DoE Joint Dark Energy Mission (JDEM) include major components of gravitational lensing surveys \cite{snap,destiny}. At the time
of writing, the JDEM selection procedure is on hold, pending the outcome of the US decadal review. 
The European Space Agency's Euclid mission concept has been designed explicitly with gravitational lensing as a 
primary science and instrumentation driver. Euclid 
will image the entire extra-galactic sky ($20,000$ square degrees) in optical and infra-red with an 
image resolution and depth similar to 
the HST Deep Field, and obtain spectra for a quarter of the lensed galaxies \cite{euclid}. The 
Euclid dark matter science objectives include  
providing a dark matter map of the Universe to a redshift of $2$, detecting hundreds of thousands of dark matter 
clusters and haloes from $10^8$~M$_{\odot}$ to $10^{15}$~M$_{\odot}$ and constraining the neutrino mass, number 
and hierarchy to percent accuracy. 

High resolution space-based imaging has also proven essential for measurements of strong gravitational lensing, in
which the most important factor is generally the number of lensed sources resolved around each foreground mass
(which can then be cross-compared to remove degeneracies), rather than the total number of lenses on the sky.
About 1 in 200 elliptical galaxies are strong lenses, and about 1 in 90
strong lenses provide something close to a double Einstein ring. An all-sky survey with Euclid should therefore
find several tens of  thousands of strong lenses, several hundred double Einstein rings, and maybe a ``golden
lens'' with more. Following up all these detections with spectroscopic observations to obtain their
redshifts will require a significant investment of time on ground-based telescopes such as some proposed 
30m-class telescopes. The proposed Observatory for Multi-Epoch Gravitational lens Astrophysics
(OMEGA) space telescope will continually monitor $\sim100$ strongly lensed Active Galactic Nuclei (AGN) using
near-UV to near-IR imaging and spectroscopy with a cadence of hours to days \cite{omega}. AGN take on different physical sizes
when observed at different wavelengths, so each imaging band will be sensitive to millilensing by different masses of
substructure in the lens. The regular imaging will calibrate the physical size of
the sources as a function of wavelength, via the ``reverberation mapping'' technique, and also measure the time 
delays between multiple images in order to better understand the lens geometry.

\subsection{Software}

Fully exploiting the wealth of new survey data will require simultaneous advances in analysis techniques,
computer hardware to implement those techniques on huge data sets, and
theoretical calculations against which the observations can be compared. As appropriate for a mature
scientific field, statistical weak lensing experiments will need to simulate the entire experiment, from the growth of
structure through observation and data processing, in order to avoid systematics and correctly interpret the
results. This begins with computer simulations of structure formation -- and while there is a substantial existing
industry of dark matter simulations, these tend to include neither the complicating baryonic processes, nor
the specific visualisation of dark matter as seen by a lensing survey. For example, $\sim10\%$ biases can be
introduced by simulations too small to include even the largest coherent structures at any redshift
\cite{fosalba08}, or by simply integrating the mass distribution along radial lines away from the Earth
\cite{vanwaerbe01}. Instead, it is necessary to raytrace backwards along the paths followed by light and
follow multiple deflections \cite{teysseier09,hilbert09}. Implementing this in petabyte-scale simulations
requires preplanned organisation of the output during runtime.


Simulations will then be required to realise images of the Universe through the Earth's atmosphere, the telescope's 
point spread function (PSF) and distortions in the optical path, and pixellisation and other effects at the detector.
All of these change the shape of a galaxy. But, while this forward process is easily 
mimicked, it is a challenging inverse problem to start
from the observed data and extract the underlying weak lensing signal. 
Many image analysis techniques have been developed to measure galaxy shapes and correct them for these effects. 
To make the challenge more difficult
still, the most distant galaxies, which contain the
largest integrated lensing signal, are very faint and noisy. 


The archetypal weak lensing measurement method KSB measures galaxies'
quadrupole shape moments to infer the shear \cite{ksb95}. 
However, the physical simplicity of KSB is known to be biased in some observational
regimes and at low signal to noise. A host of new methods have
been developed and are now being used for the most recent analyses. 
Two notable methods, {\it lensfit} \cite{lensfit1,lensfit2} and {\it shapelets}
\cite{shapelets2,shapelets3,bj02} both work by forward-fitting a model galaxy shape that has been
pre-convolved with the PSF, in order to obtain a deconvolved image. {\it Shapelets} uses a model that
generalises KSB's quadrupole moments to higher order. {\it Lensfit} uses a parametric model based on the
typical shapes of galaxies and in particular deals with the noisy regime via a fully Bayesian framework. 
The Shear TEsting Programme (STEP) \cite{heymans06,massey07d}, an
international collaboration of lensing groups, then the GRaviational lEnsing Accuracy Testing (GREAT)
\cite{great08,great10} scheme, which also included input from computer scientists and experts in machine
learning, have provided lensing simulations upon which methods can be tested. 
Both schemes took the format of a blind competition, in which simulated astronomical images
containing a known weak lensing signal were distributed to entrants (who were not told the input signal) for
independent analysis.
Flexion measurement methods are still in relative infancy, but shapelets contain all the  higher order shape
information necessary to construct flexion estimators \cite{goldberg05,massey07c,goldberg07}. The similar
{\it HOLICS} estimator \cite{okura07} is identical to shapelets (and KSB) at low order, but uses
non-orthogonal moments at higher order. This improves the signal to noise,  but at the expense of correlated
errors. All these flexion methods have been implemented on small amounts of real data, 
and the FLexion Improvement Programme (FLIP) is underway to calibrate them via simulated images.

The final step of a data analysis pipeline is to reconstruct the distribution of mass from a gravitational lensing observable. 
Weak lensing shear and flexion are both related simply to the projected mass
in Fourier space \cite{kaiser,bernstein09}, but the conversion is non-local and also noisy. Several methods have
been developed to optimally filter the reconstruction, the most successful being maximum entropy wavelet-based
techniques \cite{starck06,starck09}. Incorporating additional constraints from strong gravitational lensing
is not so simple, and usually requires iteration of a model until it matches the observed positions, fluxes,
redshifts and complex shapes of observed sources \cite{bradac04,bradac06}.
Strong lensing observations yield very high information
density along lines of sight that lead to an image, but zero
elsewhere, and two styles have been developed to realise a continuous mass model from this.
Parametric models assume the mass distribution to have some symmetry or to follow the distribution
of light within a lens but, by construction, the parameters are physical quantities of direct interest
\cite{kneib96,limousin07}. Non-parametric models avoid such assumptions, but require a large number of
multiple images to provide accurate mass reconstructions on different scales \cite{broadhurst05,coe}. One new
method uses a multi-scale adaptive grid to combine advantages of both styles \cite{jullo}. 
This technique should be able to simultaneously reconstruct the mass distribution on cluster,
group and galaxy scales, and therefore provides more accurate measurements of substructure.

\section{Conclusions}\label{sec:conclusions}

During the past decade, there has been tremendous progress in the measurement of gravitational lensing, and this has contributed to key
observations of dark matter. We now know that there is a lot of dark matter, and its distribution has been mapped on a wide range of physical
scales. Dark matter is certainly not baryonic, and measurements are beginning to ascertain properties including its dynamical temperature and
interaction cross-section through fundamental forces with both itself and baryonic matter. 

In this revolution, the techniques of gravitational lensing have moved rapidly into mainstream astronomy. Progress is most apparent in the field
of weak gravitational lensing, which is now a standard tool for many astrophysicists. The excited aniticipation of a 1997 review \cite{mellier99}
was quickly realised in the almost immediate detection of cosmic shear, and has been brought by continual observational and technical advances
into a mature scientific field.

The future prospects are bright. The first telescopes purpose-built for gravitational lensing measurements have just begun dedicated surveys.
More facilities are being planned and built -- including pioneering space-based telescopes that will provide the most discriminating data to
investigate the dark sector. To keep pace with these surveys and interpret the observations, image analysis techniques and theoretical
calculations are entering a new regime in precision. Over the next decade, particle physics may find particles that are a candidate for dark
matter; astrophysics will study the missing matter directly. Together, we tiny specs of baryonic matter may be able to join up the two sides of
the dark Universe.


\section*{Acknowledgements}

The authors would especially like to thank Alexie Leauthaud, James Taylor and Andy Green for allowing them to
include figures before their publication elsewhere. We are also grateful for invaluable discussions with Phil Marshall, Andy
Taylor, Fergus Simpson, Alina Kiessling and the organisers and participants of the 2009 W.\ M.\ Keck Institute for Space Studies workshop
``Shedding light on the nature of dark matter''. 
We would finally like to thank the two referees, whose helpful suggestions were much appreciated
and have greatly improved the whole article.
RM is supported by STFC Advanced Fellowship PP/E006450/1 and FP7 grant MIRG-CT-208994. 
TK is supported by the STFC Rolling Grant number RA0888. 
JR is supported by a EU Marie-Curie Fellowship.

\end{document}